\documentclass{aa}
\usepackage{graphicx}

\def\ltsim{\raise 2pt \hbox {$<$} \kern-1.1em \lower 4pt \hbox {$\sim$}}
\def\ltapprox{\raise 2pt \hbox {$<$} \kern-1.1em \lower 5pt \hbox {$\approx
$}}
\def\gtsim{\raise 2pt \hbox {$>$} \kern-1.1em \lower 4pt \hbox {$\sim$}}
\def\gtapprox{\raise 2pt \hbox {$>$} \kern-1.1em \lower 5pt \hbox {$\approx
$}}

\begin{document}

\title{Low frequency follow up of radio halos and relics in the GMRT Radio Halo
Cluster Survey}
\author{T.~Venturi\inst{1}, S.~Giacintucci\inst{2,3}, D. Dallacasa\inst{4},
R. Cassano\inst{1}, G. Brunetti\inst{1}, G. Macario\inst{5}, 
R. Athreya\inst{6}}
\institute
{
INAF -- Istituto di Radioastronomia, via Gobetti 101, I-40129, Bologna, Italy 
\and
Department of Astronomy, University of Maryland, College Park, MD
20742-2421
\and
Joint Space-Science Institute, University of Maryland, College Park,
MD, 20742-2421, USA
\and
Dipartimento di Astronomia, Universit\`a di Bologna, via Ranzani 1, 
I-40126 Bologna, Italy
\and
Laboratoire Lagrange, UMR7293, Universit\'e de Nice Sophia-Antipolis,
CNRS, Observatoire de la C\^ote d'Azur, 06300 Nice, France
\and
Indian Institute of Science Education and Research (IISER), Pune, India
}

\date{Received 00 - 00 - 0000; accepted 00 - 00 - 0000}

\titlerunning{Low frequency follow up of the GMRT Radio Halo Survey}
\authorrunning{Venturi et al.}

\abstract
{}
{To gain futher insight on the origin of diffuse radio sources in galaxy
clusters and their
connection with cluster merger processes, we performed GMRT low 
frequency observations of the radio halos, relics and new candidates 
belonging to the GMRT Radio Halo Cluster Sample first observed at
610 MHz.
Our main aim was to investigate their observational properties and 
integrated spectrum at frequencies below 610 MHz.}
{High sensitivity imaging was performed using the GMRT at 325 MHz and 240 MHz.
The properties of the diffuse emission in each cluster were compared to our 
610 MHz images and/or literature information available at other frequencies, 
in order to derive the integrated spectra over a wide frequency range.}
{Cluster radio halos form a composite class in terms of spectral properties. 
Beyond the classical radio halos, whose spectral index $\alpha$ is in the 
range $\sim1.2\div1.3$ (S$\propto\nu^{-\alpha}$), we found sources with 
$\alpha\sim1.6\div1.9$. This result supports the idea that
the spectra of the radiating particles in radio halos is not universal,
and that inefficient mechanisms of particle acceleration are responsible 
for their origin.
We also found a variety of brightness distributions, i.e. centrally peaked 
as well as clumpy halos. Even though the thermal and relativistic plasma
tend to occupy the same cluster volume, in some cases a positional
shift between the radio and X--ray peaks of emission is evident. 
Our observations also revealed the existence of diffuse 
cluster sources which cannot be easily classified either as halos or relics.
New candidate relics were found in A\,1300 and in A\,1682, and in some
clusters ``bridges'' of radio emission have been detected, connecting the 
relic and radio halo emission.
Finally, combining our new data with literature information, we derived the 
LogL$_{\rm X}$--LogP$_{\rm 325~MHz}$ correlation for radio halos, and investigated
the possible trend of the spectral index of radio halos with the temperature 
of the intracluster medium.}
{}
\maketitle
\section{Introduction}\label{sec:intro}

Our understanding of the interplay between the thermal and non--thermal
components in clusters of galaxies, as well as their link to the processes 
leading to the formation of massive clusters in the hierarchical scenario, 
has considerably improved thanks to the high sensitivity of radio 
interferometers such as the Very Large Array (VLA) and the Giant Metrewave 
Radio Telescope (GMRT), and to the advent of the Chandra and XMM-Newton  
X--ray observatories.

Non--thermal radio emission in galaxy clusters may take the form of radio 
halos and relics, diffuse and extended sources (up and above the Mpc) of 
low surface brightness (of the order of the $\mu$Jy arcsec$^{-2}$) with no 
obvious optical counterpart, which have so far been found in $\sim$ 40--50 
clusters.
They prove the existence of relativistic particles (with Lorentz factor 
$\gamma \gg 1000$) and large scale $\mu$G magnetic fields.
Their power--law radio spectrum, with typical spectral index 
$\alpha \sim 1.2~-~1.3$\footnote{$S_{\nu}\propto \nu^{-\alpha}$, where  
$S_{\nu}$ is the flux density at the frequency $\nu$.} is the signature of a
synchrotron origin (see Ferrari et al. \cite{ferrari08}, Venturi 
\cite{venturi11} and Feretti et al. \cite{feretti12} for recent 
observational reviews).
Beyond the similar observational properties, halos and relics differ in 
the location within the cluster and in the polarization properties. 
Radio halos are centrally located, with a rather regular shape
usually coincident with the distribution of the intracluster medium
(ICM) and are unpolarized (exception 
made for MACS\,J0717.5+3745, Bonafede et al. \cite{bonafede09}), suggesting 
that the radio emission comes from a region co--spatial with the X--ray 
emitting gas.
On the other hand, relics are found in peripheral cluster regions and 
usually exhibit an elongated structure and high fractional polarization (e.g., 
Clarke \& En\ss{}lin \cite{clarke06}, van Weeren et al. \cite{vanweeren10}),
suggesting that they are located in external regions of the clusters, or
seen in projection on the cluster centre.

The longstanding key question concerns the origin of halos and relics, 
since the radiative life--time of the radiating electrons is much shorter 
than the diffusion time necessary to cover the cluster scale volumes, 
therefore some form of in--situ particle acceleration is needed to 
account for their existence (Jaffe \cite{jaffe77}; 
see also Cassano \cite{cassano09} and Brunetti \cite{brunetti11} for 
recent reviews).

All models proposed to explain the origin of relic sources 
invoke the presence of a shock in the thermal gas, possibly induced by
dynamical activity within the cluster (En\ss{}lin et al. \cite{ensslin98};
En\ss{}lin \& Gopal--Kishna \cite{ensslin01}; Hoeft \& Br\"uggen 
\cite{hoeft07}). 
So far a spatial connection between a X--ray shock, or candidate, and 
cluster--scale radio emission in the form of a distinct radio relic, 
or an edge in the radio halo structure, has been reported only for 
a handful of clusters,  i.e.  the Bullet cluster (Markevitch et al. 
\cite{markevitch02}); A\,520 (Markevitch et al. \cite{markevitch05}); 
A\,754 (Macario et al. \cite{macario11a}); 
A\,3667 (Finoguenov et al. \cite{finoguenov10}; A\,521 
(Giacintucci et al. \cite{giacintucci08} and Bourdin et al. \cite{bourdin12});
CIZA~J\,2242.8+5301  (Ogrean et al. \cite{ogrean12}); RXCJ\,1314.5--2515 
(Mazzotta et al. \cite{mazzotta11}). 
A census of the relics studied so far is given in van Weeren et al.
\cite{vanweeren11}; a review of our knowledge of shocks and of the related 
radio emission has been recently given in Markevitch (\cite{markevitch10}).

The origin of giant radio halos is more debated. A popular scenario,
the so called  ``re--acceleration model'' (a modern elaboration of the
class of the ``in--situ'' acceleration scenario, first proposed by Jaffe
\cite{jaffe77} and Roland \cite{roland81}, and quantitatively developed 
by Schlickeiser et al. \cite{schlickeiser87}), assumes that relativistic  
electrons are re--accelerated in the ICM by MHD turbulence injected in the  
cluster volume by merger events (Brunetti et al. \cite{brunetti01}; 
Petrosian \cite{petrosian01}, Fujita et al. \cite{fujita03}). 
This scenario is motivated by the observed connection
between radio halos and cluster mergers and by considerations on the 
spectra of radio halos, which appear to favour mechanisms which are not
very efficient in the acceleration of the emitting particles (see Brunetti
\cite{brunetti11} for a recent overview).
Alternative models assume that the relativistic electrons are secondary 
products of the collisions between the intergalactic cosmic rays and the 
thermal protons in the ICM (the so--called ``secondary models'', see Dennison 
\cite{dennison1980}, Blasi \& Colafrancesco \cite{blasi99}, Pfrommer \& 
En\ss{}lin \cite{pfrommer04}, Keshet \& Loeb \cite{keshet10}). 
These models are based on the argument of cosmic ray confinement in the 
intracluster medium, and on the fact that the production of secondary
electrons, at least to some extent, is unavoidable. More recently, turbulent 
re--acceleration of relativistic protons and secondary particles has also 
been investigated (Brunetti \& Blasi \cite{bb05}, 
Brunetti \& Lazarian \cite{bl11}).

The GMRT Radio Halo Survey (Venturi et al. \cite{venturi07} and 
\cite{venturi08}, V07 and V08 respectively) has provided a first insight into
the statistical properties of radio halos. It has been shown that radio halos
are a transient phenomenon, being hosted only in $\sim$30\% of massive
clusters (Cassano et al. \cite{cassano06} and \cite{cassano08}; 
Cassano \cite{cassano09}) undergoing major mergers, as quantitatively derived 
from the analysis of the X--ray emission.
Conversely, to date no halos have been detected in dynamically relaxed clusters
(Cassano et al. \cite{cassano10a}).  These results support the importance
of turbulence in re--accelerating and confining the emitting particles
in cluster mergers (Brunetti et al. \cite{brunetti09}, En\ss{}lin et al.
\cite{ensslin11}).
\\
Turbulent re--acceleration models further predict the existence of a 
composite population of radio halos, with spectral index steeper than studied 
so far at centimeter wavelentghs.
It is now clear that such sources exist: the radio halo in A\,521, with a
spectral slope $\alpha \sim 1.9$, is considered the prototype of these 
``ultra--steep spectrum radio halos'' (USSRH) (Brunetti et al. 
\cite{brunetti08} and Dallacasa et al. \cite{dallacasa09}).
The current status of our 
knowledge of the diffuse sources in galaxy clusters can be found in the 
books of recent conference proceedings on this topic (Ferrari et al. 
\cite{ferrari11},  Dwarakanath et al. \cite{dwaraka11}).

The largest majority of radio halos and relics have been detected and imaged 
only at 1.4 GHz (i.e. Bacchi et al. \cite{bacchi03}, Govoni et al. 
\cite{govoni01} and \cite{govoni04}, Giovannini et al. \cite{giovannini09}), 
and interferometric high sensitivity observations at
frequencies $\nu \le 325$~MHz are available only in a handful of cases. 
As a result, our knowledge of the properties of halos and relics and of
their spectra at frequencies below $\nu\sim$ 325 MHz is still poor and 
incomplete.
To start filling this gap, we performed  high sensitivity low 
frequency follow--up observations of all the cluster sources (radio halos 
and relics) and candidate diffuse emission detected with the GMRT at
610~MHz, as well as the known radio halos in the GMRT cluster sample still 
lacking information at frequencies below 1.4 GHz (V07 and V08).
The clusters were surveyed at 325 MHz, and some of them were followed up 
at 240 MHz as well. 

In this paper we report on the results of our 325 MHz and 240 MHz GMRT 
observations. The paper is organised as follows: in Section \ref{sec:obs} we 
present the radio observations and data reduction; in Section \ref{sec:images} 
we present the 325 MHz and 240 MHz radio analysis of the radio halos and 
diffuse sources; in Section \ref{sec:disc} we discuss our results. 
Conclusions are given in Section \ref{sec:summary}.
\\
We adopt the $\Lambda$CDM cosmology, with H$_0$=70 km 
s$^{-1}$ Mpc$^{-1}$, $\Omega_m=0.3$ and $\Omega_{\Lambda}=0.7$. 

\section{Observations and data reduction}\label{sec:obs}

\subsection{The cluster sample}

From the GMRT radio halo cluster sample (V07 and V08) we selected all 
clusters with a diffuse source, either in the form of radio halo 
or relics, and those with ``suspect'' diffuse emission, for a low frequency 
follow--up.
Table 1 provides the list of all clusters in our project. The legend to 
Column 5 is: GRH=giant radio halo\footnote{Linear size  $\ge$ 1 Mpc as defined 
in CB05, with H$_0$=50 ~km~s$^{-1}$~Mpc$^{-1}$. This size corresponds to 
$\gtsim$ 700 kpc with the cosmology assumed in this paper.}; RH=radio halo;
Rel=relic.

Four clusters in Table 1 have been published in dedicated 
works: A\,521 (Giacintucci et al. \cite {giacintucci08} and Brunetti et 
al. \cite{brunetti08}), A\,697 (Macario et al. \cite{macario10}), A\,781
(Venturi et al. \cite{venturi11a}), RXCJ\,2003.5--2323 (Giacintucci et al. 
\cite{giacintucci09}).
In this paper we present the remaining clusters, and we
incorporate all the clusters in Table 1 in the discussion.


\begin{table*}[h!]
\caption[]{Clusters observed with the GMRT}
\begin{center}
\begin{tabular}{lcccccc}
\hline\noalign{\smallskip}
Cluster name & RA &  DEC & z & Radio Source & Ref.   & scale \\ 
                      &  (h m s) &  ($^{\circ}$  $^{\prime}$
                      $^{\prime \prime}$) &  &  & & (kpc/$^{\prime \prime}$)\\
\hline\noalign{\smallskip}
A\,2744  & 00 14 18.8 & $-$30 23 00 & 0.307 & GRH & a & 4.526 \\
A\,209    & 01 31 53.0 & $-$13 36 34 & 0.206 & GRH & a & 3.377\\
A\,521    & 04 54 09.1 & $-$10 14 19 & 0.248 & GRH+Rel & b, c, d & 3.887\\
A\,697    & 08 42 53.3 & +36 20 12 & 0.282 & GRH & e & 4.265\\
A\,781    & 09 20 23.2 & +30 26 15 & 0.298 & Rel+Candidate RH & f & 4.434 \\
Z\,2661  & 09 49 57.0 & +17 08 58 & 0.383 & Candidate RH     & a & 5.231\\
A\,1300  & 11 31 56.3 & $-$19 55 37 & 0.308 & RH+Rel & a & 4.536 \\
A\,1682  & 13 06 49.7 & +46 32 59 & 0.226 & Candidate RH+Rel & a, g & 3.626\\
RXCJ\,1314.4--2515 & 13 14 28.0 & $-$25 15 41 & 0.244 & RH+Double Rel  & a, h & 3.840\\
A\,1758N& 13 32 32.1 & +50 30 37 & 0.280 & RH & a & 4.244\\
RXCJ\,2003.5--2323 & 20 03 30.4 & $-$23 23 05 & 0.317 & GRH   & i & 4.625\\
\hline\noalign{\smallskip}
\end{tabular}
\end{center}
\label{tab:clusters}
Notes: a: This paper; b: Giacintucci et al. \cite{giacintucci08}; c: 
Brunetti et al. \cite{brunetti08}; d: Dallacasa et al. \cite{dallacasa09}; 
e: Macario et al. \cite{macario10}; f: Venturi et al. \cite{venturi11a}; 
g: Venturi et al. \cite{venturi11b}; h: Mazzotta et al. \cite{mazzotta11}; 
i: Giacintucci et al. \cite{giacintucci09}.
\end{table*}



\begin{table*}[h!]
\caption[]{Details of the GMRT observations presented in this paper.}
\begin{center}
\begin{tabular}{lccccc}
\hline\noalign{\smallskip}
Cluster name &  $\nu$ &  $\Delta \nu$ & Obs. time  & HPBW, PA $^{a}$&   rms \\ 
             &  (MHz) & (MHz) & (hr)  &
	    ($^{\prime \prime} \times^{\prime \prime}$, $^{\circ}$) & (mJy b$^{-1}$)  \\
\hline\noalign{\smallskip}
A\,2744 & 325 & 32 & 8 & 15.9$\times$8.5, 38 & 0.15 \\
A\,209  & 325 & \phantom{0}32 $^{b}$& 8 & 11.0$\times$8.7, 30 & 0.12 \\
Z\,2661 & 325 & 32 &10\phantom{0} & 10.8$\times$8.4, $-83$ & 0.12 \\
A\,1300 & 325 & \phantom{0}32 $^{b}$& 8 & 14.0$\times$8.8, 8 & 0.45 \\
A\,1682 & 240 & 8 & 6 & 13.4$\times$10.1, 47 & 0.58 \\
RXCJ\,1314.4--2515 & 325 & 32 & 8 & 15.1$\times$8.0, 32 & 0.16 \\
A\,1758N& 325 & 32 & 8 & 10.5$\times$8.4, 58 & 0.10 \\
\hline\noalign{\smallskip}
\end{tabular}
\end{center}
\label{tab:obs}
Notes: $a$: Half-power beamwidth and position angle of the full array; $b$:
the observations 
were carried out using a total bandwidth of 32
MHz (USB+LSB), but only the USB  was used for the analysis.
\end{table*}


\begin{table*}[t]
\caption[]{Properties of the discrete radio sources}
\begin{center}
\begin{tabular}{lccccc}
\hline\noalign{\smallskip}
Cluster  & Source  & RA $^a$& DEC $^a$& S$_{\rm peak, 325 MHz}$ &  S$_{\rm tot, 325 MHz}$ \\
name    &              & (h,m,s) & ($^{\deg}, ^{\prime}, ^{\prime \prime}$) &
(mJy beam$^{-1}$)  & (mJy)  \\
\noalign{\smallskip}
\hline\noalign{\smallskip}
A\,2744  & S1   & 00 14 21.6 & $-$30 25 56 & 7.4 & 14.0 \\
               & S2   & 00 14 33.8 &  $-$30 22 31 & 6.3 & 7.5 \\
               & S3   & 00 14 39.9 & $-$30 28 20  & 2.1 & 2.4 \\
               & S4   & 00 14 44.5 & $-$30 26 20  & 9.8 $^{b}$ & 83.6
               $^{b}$ \\
&&&&&\\
A\,209    & S1 &  01 31 52.5 & $-$13 37 00 & \phantom{0}40.9 $^b$ & \phantom{0}81.7$\pm$4.1 $^b$\\
               & S2 &  01 31 56.5 & $-$13 35 17 & 7.0 & 13.2$\pm$0.7 \\ 
&&&&&\\
Z\,2661    & S1 & 09 49 47.1 & +17 06 00 & \phantom{0}2.0 $^b$ & 7.8 $^b$ \\
                 & S2 & 09 49 49.1 & +17 06 57 & 3.0 & 5.1 \\
                 & S3 & 09 49 50.3 & 17 06 34 & \phantom{0}1.5 $^b$& 3.2 $^b$\\
                 & S4 & 09 49 52.5 & 17 07 34 & 6.2 & 8.7 \\ 
&&&&&\\
A\,1300   & A1  & 11 31 54.4 & $-$19 55 39 & 50.3 & 60.8 \\
                & A2  & 11 31 54.3 & $-$19 53 55 & \phantom{0}92.5 $^b$& 197.8 (216.1) $^{b,c}$\\
                & A3  & 11 31 54.9 & $-$19 52 05 & \phantom{0}14.1 $^{b}$ & 23.1  $^{b}$ \\
                & B1  & 11 31 43.7 & $-$19 52 55 & \phantom{0}19.6 $^{b}$  & 40.7
                $^{b}$  \\
&&&&&\\
A\,1682 & S1 & 13 06 45.6 & +46 33 32 & \phantom{0}400.0 $^{b}$& 1131.6 $^{b,d}$ \\
              &S2 & 13 06 49.9 & +46 33 33& 29.8 & 50.3\\  
              & S3 & 13 07 03.7 & +46 33 47 & 5.6 & 6.9 \\
&&&&&\\
RXCJ\,1314.4--2515 & S1 & 13 14 30.4 & -25 17 15 & 5.8 & 11.5 \\
                                  & S2 & 13 14 48.5 & -25 16 17 & 7.9
                                  & 18.0 \\
&&&&&\\                                
A\,1758N & S1  &   13 32 35.7 &+50 32 35 & 1.0 & 2.0 \\
                 & S2  & 13 32 39.0 & +50 33 36 & 16.9 & 21.0 \\
                 & S3 & 13 32 39.9 & +50 34 32 & 18.8& 27.2\\
                 & S4$^a$ & 13 32 54.8 & +50 31 35 & \phantom{0}145.3 $^b$ & \phantom{0}514.6 \\
                 & S5& 13 32 59.6 & +50 31 22 & 18.0 & 40.34 \\
\hline{\smallskip}
\end{tabular}
\end{center}
Notes: $a$: coordinates of the radio peak; $b$: measured using TVSTAT; $c$:
the value in brackets includes the {\it bridge} between A1 and A2;
$d$: this value includes the E tail.
\label{tab:sources}
\end{table*}
%
%
\begin{table*}[h!]
\caption[]{Properties of the diffuse cluster sources in the GMRT Radio Halo Survey}
\begin{center}
\begin{tabular}{lcccccc}
\hline\noalign{\smallskip}
Cluster name &  Radio source & $\nu$ & $S_{\nu}$ & P$_{\nu}$ & LLS  & HPBW\\
             &               & (MHz) & (mJy)    &  W~HZ$^{-1}$      & (Mpc) &  
($^{\prime \prime} \times^{\prime \prime}$) \\
\hline\noalign{\smallskip}
A\,2744 & GRH &325 & 323$\pm$26 & 9.68$\times10^{25}$ & $\sim$1.9 & 35.0$\times$35.0\\
        & Rel &325 & 122$\pm$10 & 3.66$\times10^{25}$ & $\sim$1.3 & 35.0$\times$35.0\\ 
A\,209  & GRH &325 & 74$\pm6^a$ & 8.91$\times10^{24}$ & $\sim0.6^a$& 25.0$\times$25.0\\
A\,1300 & GRH &325 & 130$\pm$10 & 3.92$\times10^{25}$ & $\sim$0.9 & 28.0$\times$28.0\\
        & Rel &325 &  75$\pm$6  & 2.26$\times10^{25}$ & $\sim$0.5 & 28.0$\times$28.0\\
   & Cand. Rel&325 &  23$\pm$2  & 6.94$\times10^{24}$ & $\sim$0.7 & 28.0$\times$28.0\\
A\,1682  & Diffuse Comp.  & 240 & 46$\pm$4 & 6.82$\times10^{24}$   & $\sim$0.3 & 18.3$\times$14.0\\
RXCJ\,1314.4--2515 & W Rel & 325 & 137$\pm$11& 2.42$\times10^{25}$ &           &26.1$\times$10.2\\
                   & E Rel & 325 &  52$\pm$4 & 9.18$\times10^{24}$ &         &26.1$\times$10.2\\
                   & RH    & 325 & 40$\pm$3  & 7.06$\times10^{24}$ &         &26.1$\times$10.2\\
A\,1758N& GRH &325 & 155$\pm$12              & 3.75$\times10^{25}$ & $\sim$1.5 &35.0$\times$35.0\\
\hline\noalign{\smallskip}
\hline\noalign{\smallskip}
A\,521$^b$  & GRH  & 240 &  152$\pm15$ & 2.79$\times10^{25}$ & $\sim$0.9& 35.0$\times$35.0 \\
A\,521$^b$  & Rel  & 240 &  180$\pm$10 & 3.30$\times10^{25}$ & $\sim$1 & 35.0$\times$35.0 \\
A\,521$^c$  & Rel  & 325 &  114$\pm$6  & 2.09$\times10^{25}$ & $\sim$1 & 16.0$\times$13.0 \\
A\,697$^d$  & GRH  & 325 & 47.3$\pm$2.7& 1.16$\times10^{25}$ &$\sim$1.3  & 46.8$\times$41.4\\
A\,781$^e$  & Rel  & 325 & 93.3$\pm$7.5& 2.61$\times10^{25}$ &$\sim$0.8  & 40.0$\times$37.0\\
RXCJ\,2003.5--2323$^f$ & GRH & 240 & 360$\pm$18 & 1.16$\times10^{26}$ &$\sim$1.4  & 35.0$\times$35.0\\
\hline\noalign{\smallskip}
\end{tabular}
\end{center}
\label{tab:obs}
Notes: $^a$: The value should be considered a lower limit, see Section 3.1.1. 
$^b$ Information taken from Brunetti et al. \cite{brunetti08}
$^c$ Information taken from Giacintucci et al. \cite{giacintucci08}. 
$^d$ Information taken from Macario et al. \cite{macario10}. 
$^e$ Information taken from Venturi et al. \cite{venturi11a}. 
$^f$ Information taken from Giacintucci et al. \cite{giacintucci09}. 
\end{table*}

\begin{table*}[h!]
\caption[]{Candidate diffuse cluster emission}
\begin{center}
\begin{tabular}{lccc}
\hline\noalign{\smallskip}
Cluster name &  Candidate source   & $\nu$ & $S_{\nu}$ \\
             &                     & (MHz) & (mJy)    \\
\hline\noalign{\smallskip}
A\,781$^a$   &   USSRH?  &  325  & $\sim$20-30 \\
A\,1682$^b$  &   GRH?    &  240  & $\sim$ 80   \\
Z\,2661$^c$  &   RH?     &  610  & $\sim$5.9  \\
\hline\noalign{\smallskip}
\end{tabular}
\end{center}
\label{tab:limits}
Notes: $^a$: From Venturi et al. \cite{venturi11a}.
$^b$: From Venturi et al. \cite{venturi11b}.
$^c$: From V08.
\end{table*}


\subsection{Data reduction and imaging}

All clusters except A\,1682 were observed at 325 MHz for a 
total of approximately 8 hours, using both the upper and lower side band 
(USB and LSB, respectively), left and right polarization, for a total 
observing bandwidth of 32 MHz. 
The data were recorded in spectral--line mode with 128 channels/band,
leading to a spectral resolution of 125 kHz/channel. 
A\,1682 was observed for $\sim$ 6 hours in dual band at 240/610 MHz as part of 
a 610 MHz re--observation (610 MHz data presented in V08), with a 240 MHz
bandwidth of 8 MHz spread over 64 spectral channels.
Night observing was carried out for all sources, to minimize the effects of 
ionosphere and radio frequency interference (RFI).
Table 2 reports the observational details. 

The USB and LSB datasets were calibrated and reduced  individually  using 
the NRAO Astronomical Image Processing System (AIPS) package. 
Strong RFI affected each observation, and in some cases only one portion of 
the band was used to produce the final images (see individual comments in the 
next section and notes to Table 2). 
For this reason, beyond the standard flagging of bad baselines, antennas and 
time ranges, we carried out a very accurate inspection of each dataset 
in order to identify and remove those data affected by RFI. 
In all cases the bandpass calibration was performed  using the flux 
density calibrator. 
The calibration solutions were applied to the data by 
running the AIPS task FLGIT, which subtracts a continuum from the channels 
in the u--v plane, determined on the basis of the bandpass shape and using 
a specified set of channels. Those data whose residuals exceeded a chosen 
threshold were flagged. 
\\
Each dataset was then averaged to 6 
channels of $\sim$2 MHz each. Given the large field of view of the GMRT, 
each step of the self--calibration cycle was carried out by means of the 
wide--field imaging technique to account for the non--planar nature of the sky. 
We used 25 facets covering a total field of view of 
$\sim 2.7^{\circ}\times 2.7^{\circ}$ . 
In those cases where both USB and LSB could be used, after a 
number of phase self--calibration cycles, the final USB and LSB datasets were 
further averaged from 6 channels to 1 single channel, and then combined 
together to produce the final images. Note that bandwidth smearing is 
relevant only at the edge of the wide field of view at 325 MHz and 240 MHz, 
and does not affect the central cluster regions presented and analysed in 
this paper.

Despite the massive editing required by the RFI in some cases, the quality of 
the images is generally good; the 1$\sigma$ rms level is in the range 
0.1--0.5 mJy beam$^{-1}$. The rms values reported in Table 2 were measured on 
the full resolution images. 
We estimate that the residual amplitude errors are of the order of 
$\ltsim$ 5\%.
\\
For all clusters we produced sets of final images, from the full GMRT 
resolution ($\sim10^{\prime\prime}$ at 325 MHz, and $\sim13^{\prime\prime}$ at
240 MHz) to tapered images with typical resolution of the order of 
$\sim 35^{\prime\prime} - 40^{\prime\prime}$. 

In order to properly image the diffuse cluster radio emission, 
it is necessary to subtract the contribution (to the flux and morphology)
of the embedded discrete sources.
Thanks to its configuration, which provides a simultanous good coverage of 
short and long spacings in the u--v plane, the GMRT is particularly
suited to perform accurate source subtraction. Where necessary, we carefully 
subtracted  the individual sources from the u--v data, and then
produced images of the diffuse cluster sources from the subtracted dataset. 
In particular, point--like sources with peak flux density $\ge5-6\sigma$ 
were subtracted from the diffuse emission in the u--v plane. 
For careful subtraction of the extended radio galaxies we made an image 
cutting the innermost regions of the u--v plane 
to avoid subtraction of any possible contribution from the diffuse cluster 
emission, and tapering the visibilities for best imaging of their extended 
emission. 
We then subtracted the clean components from the u--v data and finally 
imaged the source subtracted dataset.
For each source 
(point--like and extended), we ensured consistency between the total flux 
density subtracted as clean components from the u--v data and the flux 
density measured in the full resolution image.
\\
We are confident that any residual contribution of emission from extended
sources to the flux and morphology of the diffuse cluster emission is 
neglibible.
The contribution 
of faint embedded sources whose flux density is just below our threshold is 
also negligible: radio source counts show that the number of radio sources with 
mJy and sub--mJy flux density is of the order of $\sim$0.01 for a sky 
region with average largest angular size of $\sim 4-5^{\prime}$, as is the 
case for the cluster sources presented in this paper 
(i.e.  Bondi et al. \cite{bondi07}, Venturi et al. \cite{venturi02} and 
references therein).

Considering the uncertainties in the source subtraction procedure, we
associate a final error of 8\% to the flux density of the diffuse sources.

\section{Diffuse cluster sources at low frequency}\label{sec:images}

For each cluster presented in this Section, Table 3 provides the
position and flux density of the individual radio sources embedded in
the diffuse emission, as measured on the full resolution images with the 
AIPS task JMFIT for unresolved sources and TVSTAT for extended sources.
Table 4 summarizes the observational properties of the diffuse emission 
in each cluster (the morphological classification is given in Col. 2, the 
total flux density and linear size are reported in Col. 4 and 5 respectively) 
imaged after subtraction of the discrete radio galaxies listed in 
Table 3.
The bottom part of Table 4 reports the information for the clusters
presented in separate papers. In Table 5 we list the candidate diffuse
cluster sources which still need to be confirmed.

\subsection{Clusters with a giant radio halo}

\subsubsection{A\,209}

\begin{figure*}[htbp]
\centering
\includegraphics[angle=0, scale=0.79]{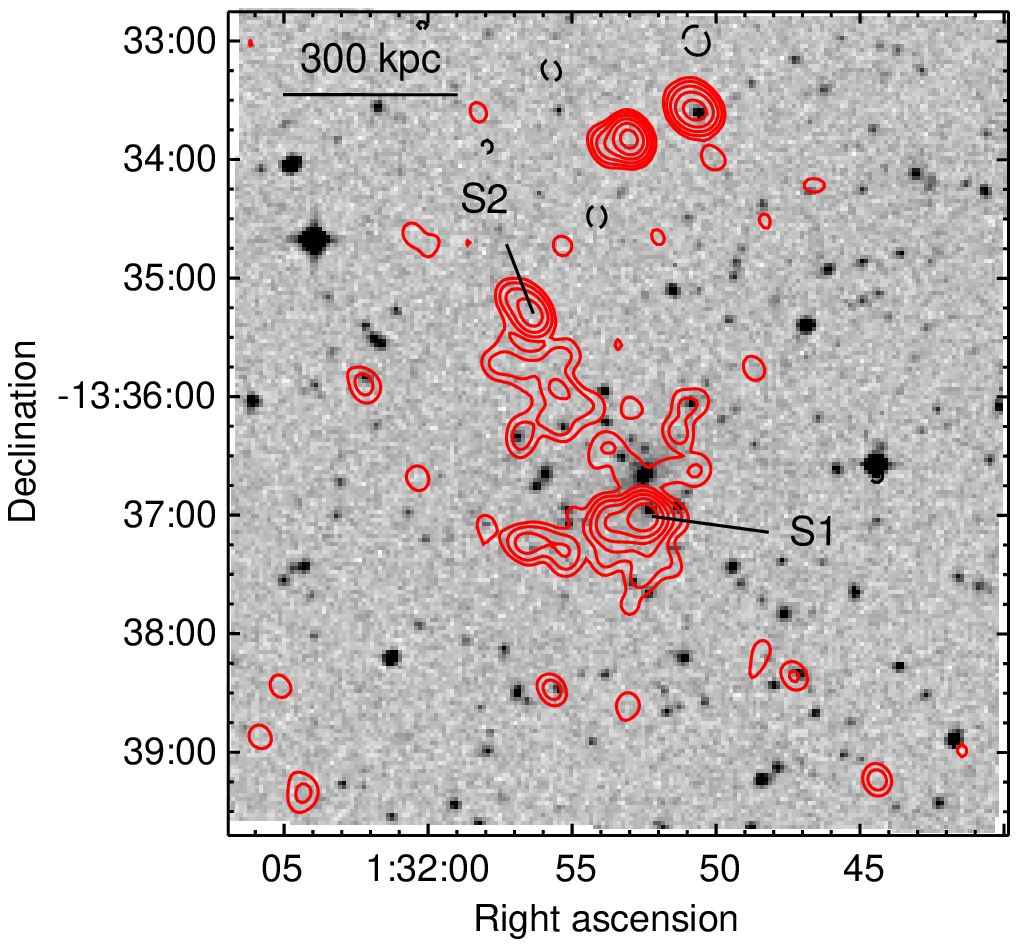}
\hspace{0.5cm}
\includegraphics[angle=0,scale=1.18]{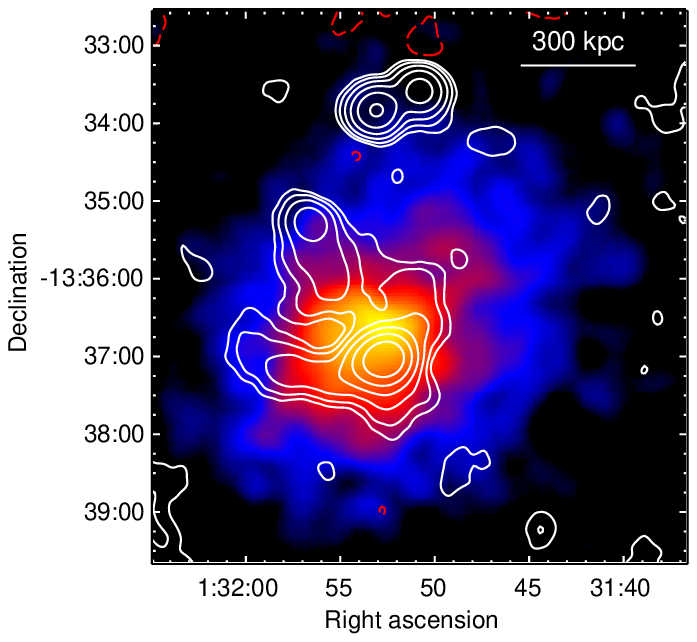}
\caption{{\it Left:} GMRT 325 MHz radio contours of A\,209 at full resolution 
($11.0^{\prime \prime}\times 8.7^{\prime \prime}$, p.a. 30$^{\circ}$), overlaid
on the optical POSS--2 image. The 1$\sigma$ level in the radio image is 
0.12 mJy beam$^{-1}$ and red contours are spaced by a factor 2 starting from 
+3$\sigma=0.36$ mJy beam$^{-1}$. Black dashed contours correspond to the 
$-3\sigma$ level. Labels S1 and S2 show are the radio galaxies embedded
in the radio halo emission. {\it Right:}
GMRT 325 MHz radio contours at the resolution of 
$25.0^{\prime \prime}\times 25.0^{\prime \prime}$, p.a. 0$^{\circ}$,
superposed to the smoothed Chandra X-ray image. No source subtraction
has been performed. White contours start at $+3\sigma=0.9$ mJy beam$^{-1}$ 
and then scale by a factor 2. The $-3\sigma$ level is shown as red dashed 
contours.} 
\label{fig:a209}
\end{figure*}

A\,209 (z=0.206) is a massive merging cluster with 
$L_{\rm X, [0.1-2.4~keV]}=6.3\times10^{44}$ erg s$^{-1}$ 
(see V07 for a review of its properties). It hosts a giant radio halo, 
imaged with the GMRT at 610 MHz (V07) and with the VLA at 1.4 GHz 
(Giovannini et al. \cite{giovannini09}, Giacintucci et al. \cite{giacintucci13},
hereinafter G12).
\\
The 325 MHz data were affected by strong interference, which required 
massive data editing and prevented high quality imaging of the diffuse 
cluster emission: no individual source subtraction was performed for
this cluster. Our images, presented in Fig. \ref{fig:a209}, 
are in agreement with those at 610 MHz, both in shape and extent (Fig. 2
in V07).
The radio halo covers only the most luminous part of the X--ray emission,
and is smaller than imaged at 1.4 GHz, most likely as consequence of the
RFI removal editing, which forced us to drop many short spacings.
The flux density of the radio halo, measured from the image shown in the
right panel of Fig. \ref{fig:a209} after subtraction of the flux
density of the discrete radio sources S1 and S2 reported in Table 3
(Fig. \ref{fig:a209} left), is 
$S_{\rm 325~MHz}=74\pm6$ mJy. We consider this value, as well as the 
largest linear size measured from the right panel of Fig. \ref{fig:a209}
(Table 4), as lower limits. V07 reported a flux density 
$S_{\rm 610 MHz} = 24.0 \pm 3.6$; Giovannini et al. (\cite{giovannini09})
report a value at 1.4 GHz of $S_{\rm 1.4 GHz} = 16.9$ mJy, while a 
re--analysis of the archival VLA 1.4 GHz data performed in G12 led to
a flux density estimate measurement $S_{\rm 1.4 GHz} = 15.0\pm0.7$ mJy.
Considering that 
the flux density of the radio halo at 610 MHz is most likely understimated 
(as discussed in V08), and in the light of the limited quality of the 325 MHz 
data, we could not perform a spectral study for this cluster and did
not include A\,209 in the discussion carried out in Sect. 4.4.1.

\subsubsection{A\,1758N}

A\,1758N (z=0.2782, Boschin et al. \cite{boschin12};
$L_{\rm X, [0.1-2.4 KeV]}=12.3\times10^{44}$ erg
s$^{-1}$) is part of a well--known pair of two merging clusters, A\,1758N and 
A\,1758S, separated by $\sim~8^{\prime}$ in the plane of the sky and 
$\sim~2100$ km s$^{-1}$ in the velocity space, both formed by two
merging clusters (Ragozzine et al. \cite{ragozzine12}).
A detailed Chandra and 
XMM--Newton X--ray study (David \& Kempner \cite{DK04}) reveals that both 
clusters are characterised by very complex X--ray morphology, with several 
clumps of emission, as commonly found in cases of ongoing cluster merging 
activity. 
A\,1758N shows more extreme X--ray properties, with higher X--ray luminosity 
and temperature compared to A\,1758S. Its total mass is 
$\sim 2-3 \times 10^{15}$M$_{\odot}$, and a new optical analysis confirms
that it is a recent merger (Boschin et al. \cite{boschin12}, 
Ragozzine et al. \cite{ragozzine12}).
A\,1758N is also much more interesting in 
the radio band, with a well--known central narrow angle tail radio galaxy 
(O'Dea \& Owen \cite{O'DO85}) and diffuse central emission visible both on 
the Northern VLA Sky Survey (NVSS) and on the WEsterbork Northern Sky Survey 
(WENSS), as reported in Kempner \& Sarazin (\cite{KS01}). 
VLA observations at 1.4 GHz were recently presented in Giovannini et al. 
(\cite{giovannini09}) and re--analysed in G12.


\begin{figure*}[htbp]
\centering
\includegraphics[angle=0,scale=0.79]{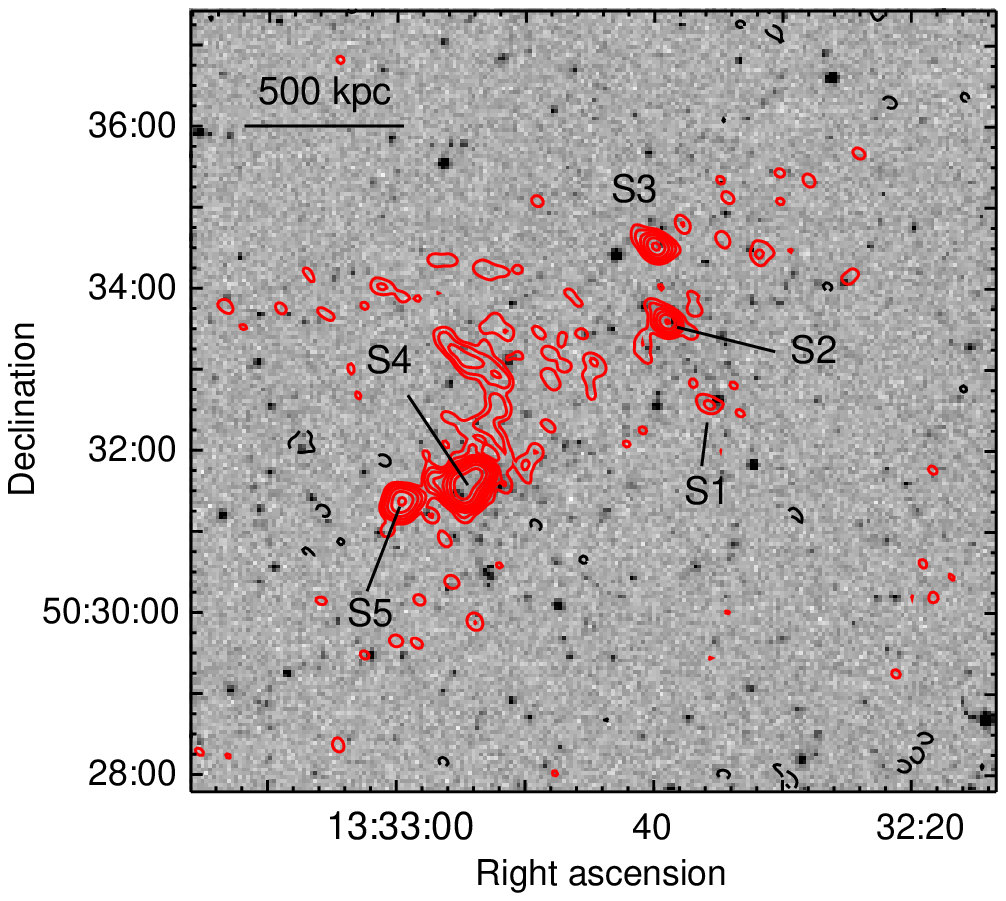}
\hspace{0.5cm}
\includegraphics[angle=0,scale=0.79]{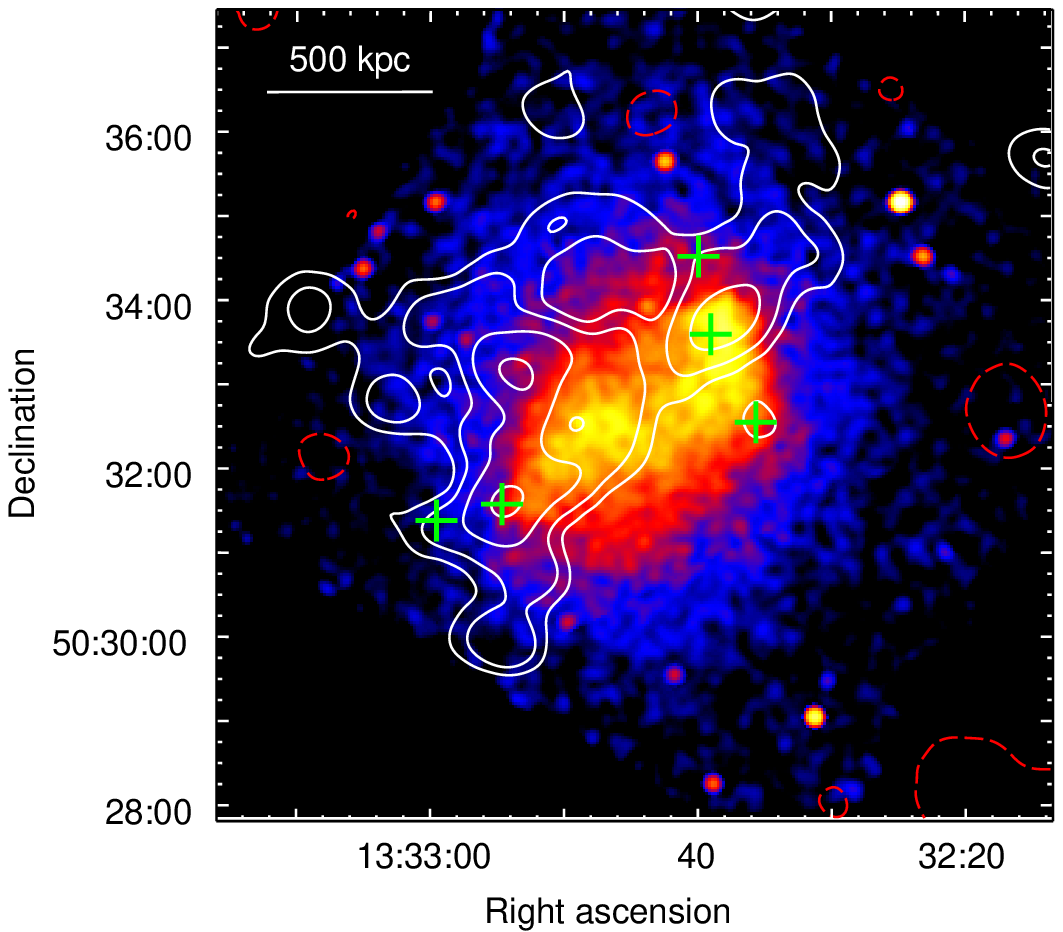}
\caption{{\it Left}: GMRT 325 MHz radio contours of the emission from A\,1758N
at the full resolution (10.5$^{\prime\prime}\times8.4^{\prime\prime}$,
p.a. 58$^{\circ}$), overlaid on the POSS--2 red image. The 1$\sigma$
noise level is 0.1 mJy beam$^{-1}$. Red contours start at 0.4 mJy beam$^{-1}$ 
and scale by a factor of 2. Black dashed contours 
correspond to the $-3\sigma$ level. Individual radio galaxies are
labelled from S1 to S5. {\it Right}: 325 MHz contours of the giant radio halo
overlaid on the smoothed Chandra X-ray image. The radio image has been 
obtained after subtraction of the discrete embedded sources (S1 to S5), whose 
position is marked as green crosses. The restoring beam is 
35.0$^{\prime\prime}\times35.0^{\prime\prime}$, p.a. 0$^{\circ}$. 
The 1$\sigma$ is 0.4 mJy beam$^{-1}$.  White contours
start at $+3\sigma=0.12$ mJy beam$^{-1}$ and scale by a factor 2. The
$-3\sigma$ level is shown as red dashed contours.}
\label{fig:a1758_tot}
\end{figure*}

The images of A\,1758N are shown in Fig. \ref{fig:a1758_tot}.
Their quality is one of the best in our GMRT 325 MHz  atlas (see Table 2). 
The left panel shows the total radio emission 
from the cluster central region imaged at full resolution overlaid on
the optical image; the right panel shows the overlay between the Chandra 
X--ray image and the diffuse emission, after subtraction of the discrete 
point--like and extended sources 
S1 to S5 from the u--v data (see Section 2.2, Fig. \ref{fig:a1758_tot}
the left panel, and Table 3).
The diffuse radio emission covers only a portion of the X--ray brightness
distribution. It has a filamentary structure, and is aligned in the 
North--West/South--East direction. The radio and X--ray peaks of
emission are  in good spatial coincidence in the north--western part of the 
cluster, while a misplacement is clear in the south--eastern cluster region.

The total flux density of the diffuse emission, measured within the 3$\sigma$ 
contour, is S$_{\rm 325~MHz}=155\pm12$ mJy. 
The total size of this structure is $\sim$ 1.5 Mpc.
The overall shape and size of the diffuse emission detected at 325 MHz is in 
reasonable agreement with the re--analysis carried out in G12. Giovannini 
et al. (\cite{giovannini09}) classify the radio emission as a radio halo 
and a double relic. Indeed the morphology of this source is very complex,
however we find it difficult to disentangle possible different components.
The overall spatial coincidence between the radio and X--ray emission,
as clear from the right panel of Fig. \ref{fig:a1758_tot}, suggests that the 
source is centrally located. Hence, we consider it a giant radio halo. 
\\
The spectral index of the whole emission, using the total 1.4 GHz flux density 
reported in G12 (S$_{\rm 1.4~GHz}=23\pm5$ mJy, integrated over the same
area) is $\alpha_{\rm 325~MHz}^{\rm 1.4~GHz}=1.31\pm0.16$. 

\subsection{Clusters with halo and relics}

Three clusters in the sample host multiple diffuse cluster sources.
A\,1300 and A\,2744, both located at z$\sim$0.3 and with similar X--ray 
luminosity, host a giant radio halo and a relic. They belong to the
GMRT radio halo cluster sample, but being known radio halos they were
not observed at 610 MHz in our earlier works (V07 \& V08).
RXCJ\,1314.5--2515 hosts spectacular diffuse emission in the shape of
a radio halo and two very long and symmetric relics.

\begin{figure*}[htbp]
\centering
\includegraphics[angle=0,scale=0.88]{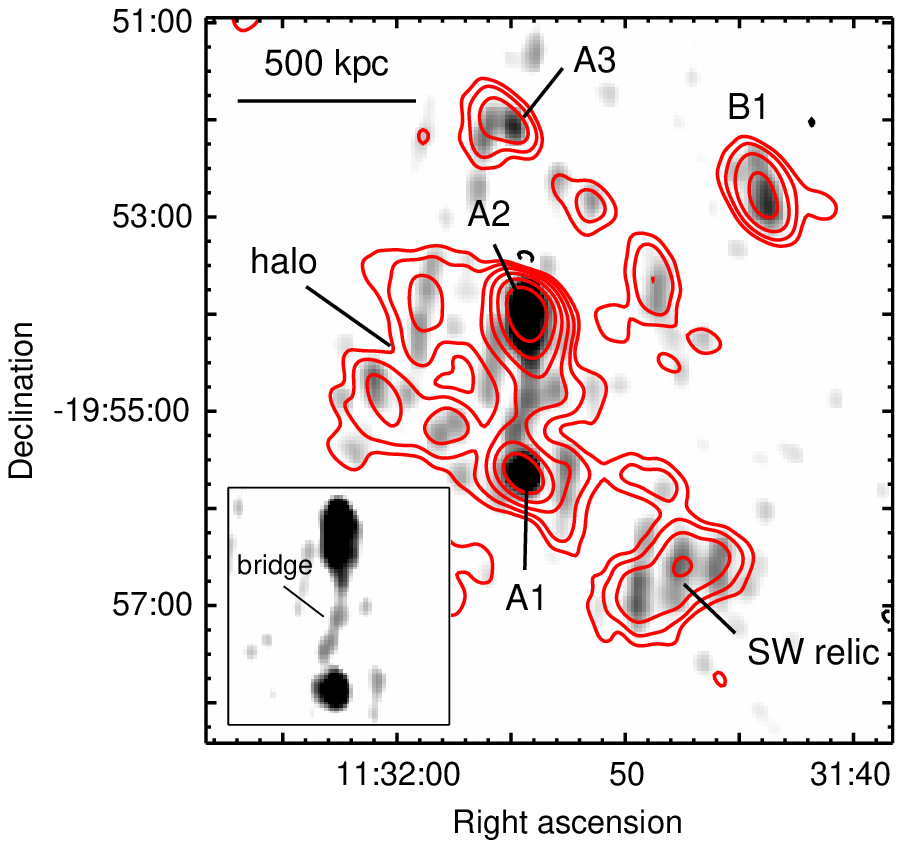}
\hspace{0.5cm}
\includegraphics[angle=0,scale=0.71]{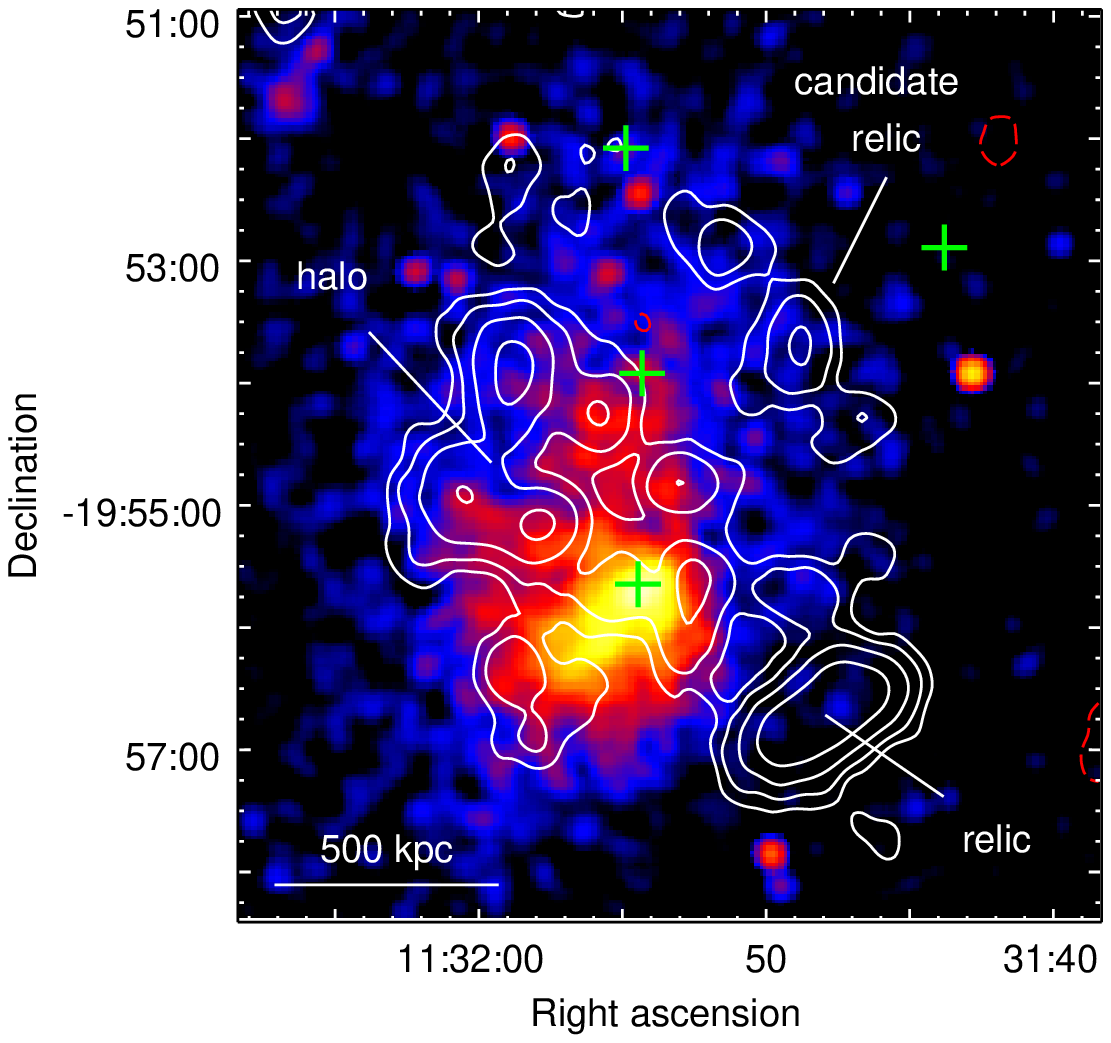}
\caption{{\it Left:} GMRT 325 MHz radio contours of the emission from A\,1300
at the resolution of 27.6$^{\prime\prime}\times18.1^{\prime\prime}$,
p.a. 43$^{\circ}$, overlaid on the grey scale full resolution image 
(14.0$^{\prime\prime}\times 8.8^{\prime\prime}$, p.a. 8$^{\circ}$). 
The 1$\sigma$ noise level of the contour image is 0.6 mJy beam$^{-1}$.
Red contours start at 1.8 mJy b$^{-1}$ and then scale by a factor of 2. 
The 1$\sigma$ 
noise level of the grey scale image is 0.45 mJy beam$^{-1}$. Individual radio 
galaxies are labelled following the notation in R99. 
{\it Right:} Radio contours of the central cluster region after subtraction 
of the discrete radio sources (the green crosses mark their location).
The restoring beam is 
28.0$^{\prime\prime}\times28.0^{\prime\prime}$, p.a. 0$^{\circ}$. 
The 1$\sigma$ is 0.5 mJy beam$^{-1}$. White contours start at 
1.5 mJy beam$^{-1}$ and then scale by a factor of 2. The $-3\sigma$ level is 
shown as red dashed contours.
}
\label{fig:a1300_tot}
\end{figure*}

\subsubsection{A\,1300}

A\,1300 is located at z=0.308 and has 
L$_{\rm X, [0.1-2.4KeV]}=1.4\times10^{45}$ erg s$^{-1}$. It hosts a radio halo,
first reported in Reid et al. (\cite{reid99}, hereinafter R99) 
on the basis of radio observations carried out with the Molonglo 
Observatory Synthesis Telescope (MOST) at 843 MHz and with the Australia
Telescope Compact Array (ATCA) from 1.34 GHz to 8.6 GHz.
With a temperature of $\sim~11$ keV (Pierre et al. 1997), A\,1300
is one of the hottest clusters known. 
Optical and X--ray analysis (Pierre et al. \cite{pierre97}, Lemonon et al. 
\cite{lemonon97}, Ziparo et al. \cite{ziparo12}) suggest that it is a 
post--merger.
The radio observations by R99 show that A\,1300 is very active
in the radio band. Beyond the radio halo, a relic is located in the 
south--western periphery of the cluster, and a number of radio galaxies with
extended emission are found at the cluster centre. In a recent optical
and X--ray analysis, Ziparo et al. (\cite{ziparo12}), reported on a possible
shock front (consistent with a Mach number $M$=1.2$\pm$0.1) and on a sharp edge
in the galaxy distribution at the location of the relic.

Due to strong interference, only the upper side band of the GMRT 325 MHz 
dataset could be used in our imaging process. 
Fig. \ref{fig:a1300_tot} (left) shows the radio emission from the cluster 
centre (contours) at the intermediate resolution of 
26.7$^{\prime\prime}\times18.1^{\prime\prime}$, 
overlaid on the full resolution image 
(14.0$^{\prime\prime}\times 8.8^{\prime\prime}$), displayed in grey scale to 
highligt the position and morphology of the radio galaxies embedded in the 
diffuse cluster emission. 
The individual radio sources are labelled following the notation in R99 and
their flux density is reported in Table 3. An elongated, thin
structure, hereinafter referred as {\it bridge} (see inset in
Fig. \ref{fig:a1300_tot}) connects the head--tail radio galaxy A2 to the point 
source A1. It is not clear whether such feature is an extension of the tail of 
A2 or a brightness peak of the diffuse radio halo.
For this reason, Table 3 reports the flux density of A2 with and without 
the contribution of the {\it bridge}, which accounts for $\sim$18 mJy. 
Note that the radio sources labelled B2, B3 and 12 in R99 are undetected in 
our 325 MHz images, most likely due to their spectral slope.
We subtracted the components of the individual radio galaxies (Table
3) from the u--v data (we considered the {\it bridge} as part of A2 and we
subtracted it as well) and produced a low resolution image 
of the residual diffuse emission, which is shown in the right panel
of Fig. \ref{fig:a1300_tot}, overlaid on the smoothed Chandra X--ray image. 
The 325 MHz brightness distribution of the radio halo is fairly uniform.
 
The shape of the radio halo at this frequency is similar to the higher 
frequency images, with the diffuse emission extending mainly East of the 
two radio sources A1 and A2. The overall size is also comparable to the one 
reported in R99, with a largest linear size (LAS) of $\sim$3.3$^{\prime}$ 
(largest linear size LLS$\sim890$ kpc). 
The S--W relic (B3 in R99), too, is similar in shape, extent 
and brightness distribution to the earlier images, and its LLS at 325 MHz
is $\sim 450$ kpc.
Our image shows the existence of an arc--shaped elongated feature, not
visible in any of the images published in R99, located
North--West of the cluster centre, which accounts for 
$S_{\rm 325 MHz}=23\pm2$  mJy and is $\sim700$ kpc in extent.
The overall morphology and location suggest
that it might be another relic (see Fig. \ref{fig:a1300_tot} right panel).
\\
The total flux density measured in the radio halo, integrating 
within the 3$\sigma$ contour level in Fig. \ref{fig:a1300_tot} (right),
is $S_{\rm 325 MHz}=130\pm10$ mJy (the flux density of the {\it bridge} was
not included).
The flux density of the S--W relic is $S_{\rm 325 MHz}=75\pm6$ mJy.

We combined our flux density values with those reported by R99 in the 
frequency range 843 MHz -- 2.4 GHz to provide an estimate of the spectral index 
for the halo and the SW relic. 
The 1.36 GHz flux density of the radio halo reported in R99 is 10 mJy
(consistent with the value we measured on the NVSS after subtraction
of the contribution of the two embedded sources A1 and A2), which provides 
a spectral index $\alpha_{\rm 325~MHz}^{\rm 1.4~GHz}$=1.8. 
The inclusion of the bridge in the 325 MHz emission from the halo would
further steepen the spectrum (up to $\alpha$=1.9).
The S--W relic has $\alpha_{\rm 325~MHz}^{\rm 843~MHz}$=0.86, which steepens to 
$\alpha_{\rm 1.34~GHz}^{2.4~GHz}=1.3$. We are aware that it is impossible 
to account for the different u--v coverages of the ATCA, MOST, GMRT and VLA. 
For this reason we regard these spectral considerations as indicative and
we do not include this cluster in the discussion performed in Sect. 4.4.1.

The overlay in the right panel of Fig. \ref{fig:a1300_tot} shows that the 
radio halo extends mainly North--East of the peak in the X--ray surface
brightness. Both the S--W  and the candidate relics are located at the 
border of the detected X--ray emission. 

\subsubsection{A\,2744}

A\,2744 is located at the redshift z=0.3066 and has
L$_{\rm X, [0.1-2.4 KeV]}=1.2\times10^{45}$ erg s$^{-1}$.
The cluster hosts a well known spectacular diffuse radio emission, with 
a giant radio halo and a giant relic. VLA radio observations at 1.4
GHz and 327 MHz are reported in 
Govoni et al. (\cite{govoni01}) and Orr\'u et al. (\cite{orru07}) 
respectively.
\\
The cluster is a complex galaxy merger and has been studied in detail in a 
wide range of energy bands.
Chandra observations (Kempner \& David \cite{KD04}) show that the
cluster is formed by a main irregular component and a smaller 
subcluster to the North--West. The main cluster shows very complex
substructure with a strong peak and a number of ``ridges'' in different
directions. The subcluster is connected to the main one by a bridge of
fainter X--ray emission, and its brigthtness distribution suggests that
it is moving towards the main condensation (Kempner \& David \cite{KD04}).
The very unrelaxed dynamical state of A\,2744 is confirmed by optical
spectroscopy, which reveals a redshift bimodal distribution and an
unusually high merger velocity (Girardi \& Mezzetti \cite{GM01}).
Merten et al. (\cite{merten11}) recently concluded that the main
cluster and the subcluster are both post--mergers, and show properties 
similar to the Bullet cluster, with a misplacement between the gas and
the dark matter.


\begin{figure*}[htbp]
\centering
\includegraphics[angle=0,scale=1.49]{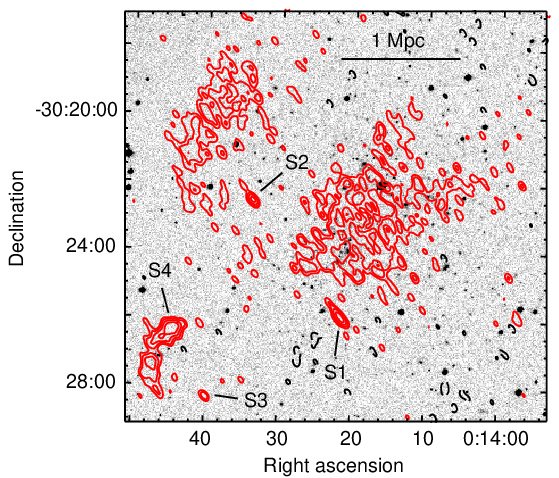}
\hspace{0.3cm}
\includegraphics[angle=0,scale=1.045]{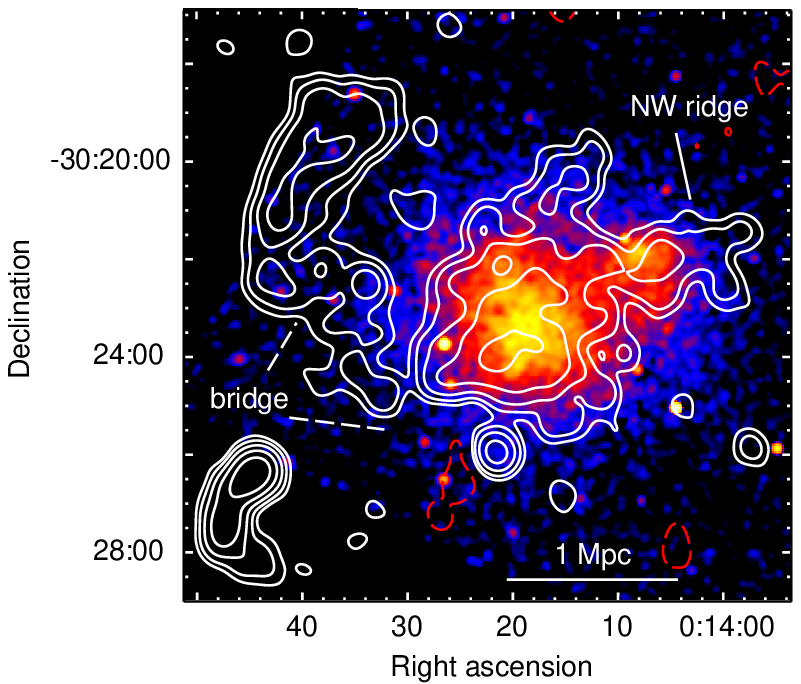}
\caption{{\it Left}: GMRT 325 MHz radio contours of A\,2744 at the full 
resolution of 15.9$^{\prime\prime}\times8.5^{\prime\prime}$,
p.a. 38$^{\circ}$, overlaid on the POSS--2 red image.
The 1$\sigma$ noise level is 0.15 mJy beam$^{-1}$. Red contours start at 0.5  
mJy beam$^{-1}$ and then scale by a factor of 2. Dashed, black contours 
correspond to the $-3\sigma$ level. Individual radio galaxies are labelled 
from S1 to S4. {\it Right}: Low resolution
(35.0$^{\prime\prime}\times35.0^{\prime\prime}$, p.a. 0$^{\circ}$)
radio contours overlaid on the smoothed Chandra X--ray image. 
The 1$\sigma$ noise level in the radio image is 0.35 mJy beam$^{-1}$.
White contours start at 1 mJy beam$^{-1}$ and then scale by a factor of 2. The
$-3\sigma$ level is shown as dashed, red contours.}
\label{fig:a2744_tot}
\end{figure*}

Fig. \ref{fig:a2744_tot} shows our 325 MHz GMRT images. Both the radio 
halo and the relic are already visible in the full resolution image (left 
panel). The two sources are connected by a faint bridge (right panel), 
visible 
also in the VLA 1.4 GHz image in Govoni et al. (\cite{govoni01}).
Thanks to the better sensitivity ($\sim$ a factor of 3) of our image 
compared to the those in Orr\'u et al. (\cite{orru07}), the radio halo in 
A\,2744 is larger than previously imaged at 325~MHz. In particular, we 
detect a north--western ridge, 
labelled in the right panel of Fig. \ref{fig:a2744_tot}, along the 
same direction of the subcluster X--ray emission. The largest linear 
size of the diffuse sources from the low resolution image (right panel) is
LLS$\sim$ 1.9 Mpc for the halo (including the NW ridge) and 
LLS$\sim$1.3 Mpc for the relic. The bridge extends for $\sim$ 700 kpc between 
the halo and relic.

The radio/X--ray overlay in Fig. \ref{fig:a2744_tot} (right) clearly shows 
that the emission from
the radio halo extends over the whole X--ray emission, and beyond, with a
positional shift between the X--ray and radio peaks both for the main cluster 
and for the N--W subcluster. The radio brightness distribution is asymmetrically
peaked, and the southern part of the radio halo is edge sharpened. 
It is noteworthy that 
Markevitch (\cite{markevitch10}) recently reported on the detection of a 
X--ray surface 
brightness edge, suggesting the presence of a shock front, at the location 
of this radio edge sharpening. The front has been confirmed by the 
Chandra analysis presented by Owers et al. (\cite{owers12}), who derived 
a Mach number of $M$=1.4 for the shock (see Fig. 12 in 
Markevitch \cite{markevitch10} for a radio/X--ray overlay of the shock region). 
Similar radio/X--ray spatial correlations have been observed in few other 
clusters with a giant radio halo, e.g., A\,520 (Markevitch et al. 2005) and 
A\,754 (Macario et al. \cite{macario11a}).
\\
As it is commonly found, the relic has a higher 
surface brightness compared to the halo, and its size is in agreement with the 
literature images. The radio/X--ray overlay confirms that the relic is located 
just outside the region where significant X--ray emission is detected
(see also Govoni et al. \cite{govoni01}). \\
From the low resolution image we measured a total flux density  
$S_{\rm 325~MHz}=323\pm26$ mJy for the halo (excluding the bridge and source 
S1), and $S_{\rm 325~MHz}=122\pm10$ mJy for the relic (excluding the bridge
and source S2). The flux density in the bridge is $\sim~30$ mJy. The
measurements were taken within the 3$\sigma$ contour level.
Both values are considerably higher than those reported in Orr\'u et al.
(\cite{orru07}). 
\\
A\,2744 is included in the re--analysis carried out in G12. Using their 
1.4 GHz flux density values over the same area, 
i.e.  S$_{\rm 1.4~GHz}=57\pm3$ and 20$\pm$1 mJy respectively for the halo and 
for the relic, we obtain $\alpha=1.19^{+0.08}_{-0.11}$ for the halo, and 
$\alpha=1.24\pm0.10$ for the relic.

\subsubsection{RXCJ\,1314.4--2515}

The cluster RXCJ\,1314.4-2515 (z=0.244,  
L$_{\rm  X, [0.1-2.4KeV]}=1.1\times10^{45}$ erg s$^{-1}$) is a spectacular 
example of a major merger with a massive subcluster (Valtchanov et al.
\cite{valtchanov02}). 
In was the first cluster where a radio halo and two relics were 
discovered
(Feretti et al. \cite{feretti05}, V07). So far only two clusters
with such features have been found (CIZA~J\,2242.8+5301, van Weeren et al. 
\cite{vanweeren11b}; MACS~J\,1752.0+4440, van Weeren et al. 
\cite{vanweeren12}).
The two relics are elongated, Mpc-size structures located in opposite 
directions with respect to the cluster centre.
Deep XMM--Newton observations led to the discovery of a prominent shock 
front, with Mach number M$\sim$ 2.5, remarkably coincident with the outer 
border of the western relic and likely produced by the merger of the 
cluster with a massive subcluster (Mazzotta et al. \cite{mazzotta11}).

\begin{figure*}[htbp]
\centering
\includegraphics[angle=0,scale=0.85]{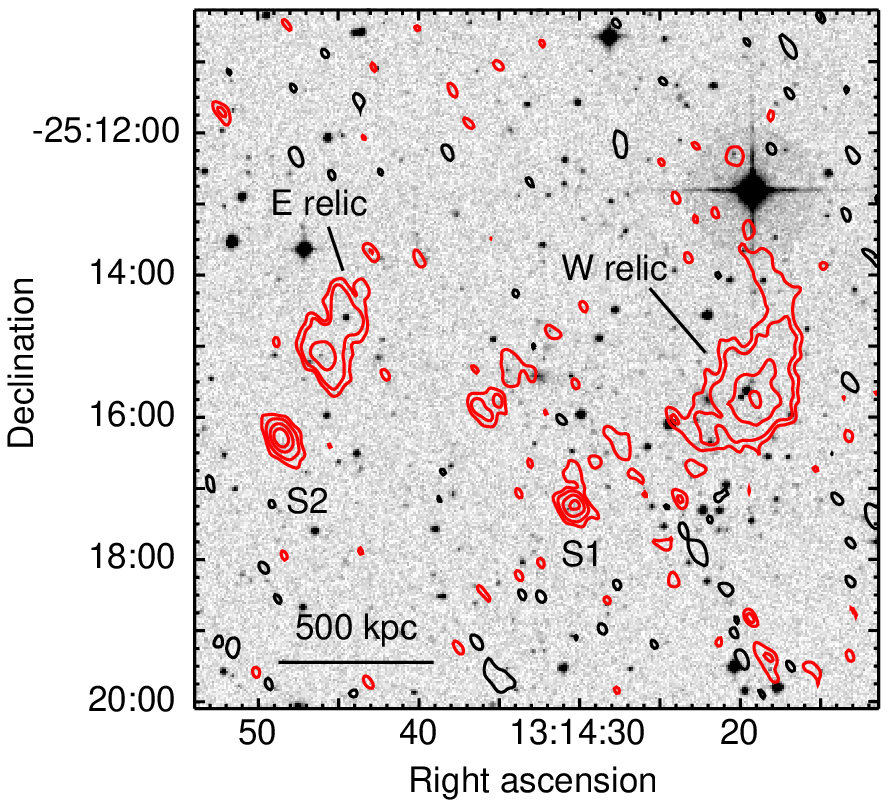}
\hspace{0.5cm}
\includegraphics[angle=0,scale=0.9]{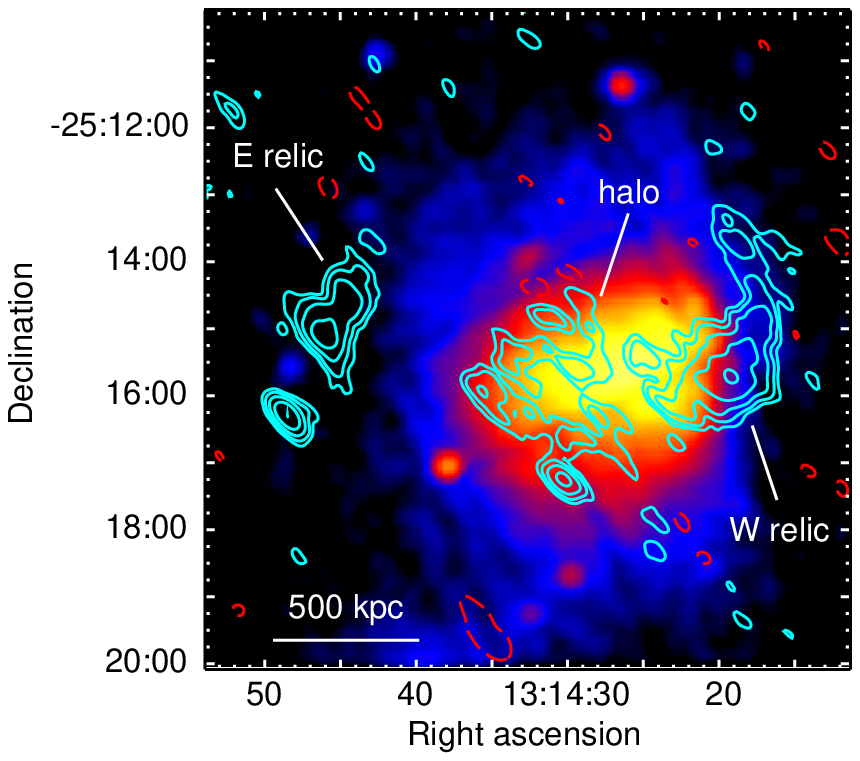}
\caption{{\it Left}: GMRT 325 MHz image of RXCJ\,1314.4--2515 at the 
resolution of 15.1$^{\prime\prime}\times8.0^{\prime\prime}$,
p.a. 32$^{\circ}$, overlaid on the POSS--2 red image.
The 1$\sigma$ noise level is 0.16 mJy beam$^{-1}$. Red contours start at 0.5  
mJy beam$^{-1}$ and then scale by a factor of 2. Black contours 
correspond to the $-3\sigma$ level. Individual radio galaxies are
labelled S1 and S2.
{\it Right:} GMRT 325 MHz radio contours overlaid on
the smoothed XMM--Newton X--ray image. The resolution of the radio 
image is 26.1$^{\prime\prime}\times10.2^{\prime\prime}$, p.a. $43^{\circ}$.
The 1$\sigma$ noise level in the radio image is $\sim$0.2 mJy
beam$^{-1}$. Cyan contours start at 0.6 mJy beam$^{-1}$ and then
scale by a factor of 2. The $-3\sigma$ level is shown as red dashed contours.}
\label{fig:rxcj1314}
\end{figure*}

Our GMRT 325 MHz observations were affected by a number of antenna failures 
and strong interference, nevertheless we managed to image the brightest 
features of the diffuse emission.
The 325 MHz cluster radio emission is reported in Fig. \ref{fig:rxcj1314}, 
where the full resolution image is overlaid on the optical frame (left) and 
a tapered image (26$^{\prime\prime}\times10^{\prime\prime}$) is overlaid
on the smoothed XMM-Newton image (right).

Due to the poor quality of our dataset, both relics are considerably smaller 
in size compared to the 610 MHz images published in V07 and to the VLA 1.4 GHz 
images in Feretti et al. (2005). In the light of all this, the flux density 
values reported in Tab. 4 should be considered as lower limits, and any 
spectral consideration is deferred to a detailed ongoing multifrequency study 
(Giacintucci et al. in progress).

\subsection{Clusters with complex diffuse emission}

\subsubsection{A\,1682: a very complex cluster}
A\,1682 (z=0.226, L$_{\rm X, [0.1-2.4 KeV]}=7.0\times10^{44}$ erg s$^{-1}$) 
is a merging
massive cluster (Morrison et al. \cite{morrison03}) with very complex
radio emission. High sensitivity and high resolution observations in the
radio band (GMRT at 610 MHz) were first published in V08, and show that 
the central compact emission visible on the NVSS is actually a blend 
of a number of features: {\it (a)} a strong radio galaxy, associated with 
the dominant m$_g$=18.0  galaxy  at the cluster centre (z=0.21839, 
Koester et al. \cite{koester07}), which diffuses into a tail; 
{\it (b)} two filamentary structures of unclear origin, which were named 
{\it S--E} and {\it N--W ridges} in V08; {\it (c)} diffuse excess emission 
over a $\sim$ Mpc--size  region, suggestive of the presence of an underlying 
giant radio halo. 

\begin{figure*}[htbp]
\centering
\includegraphics[angle=0,scale=1.1]{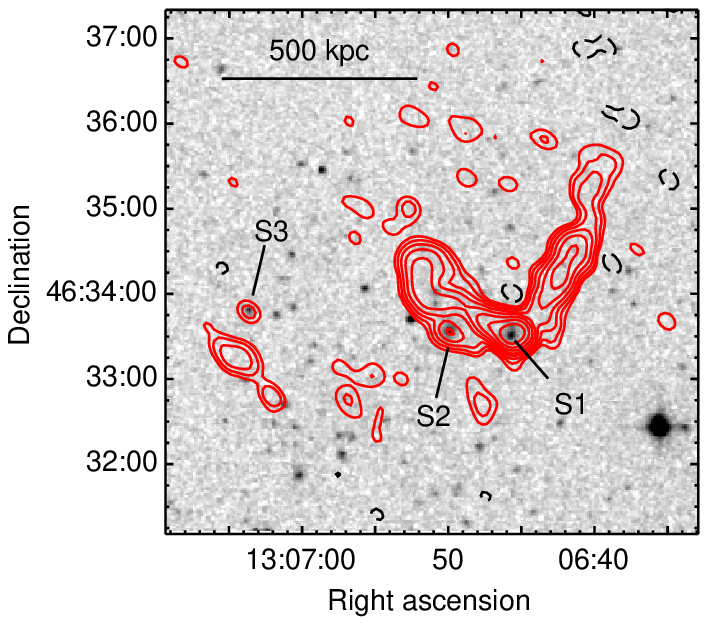}
\hspace{0.5cm}
\includegraphics[angle=0,scale=0.4]{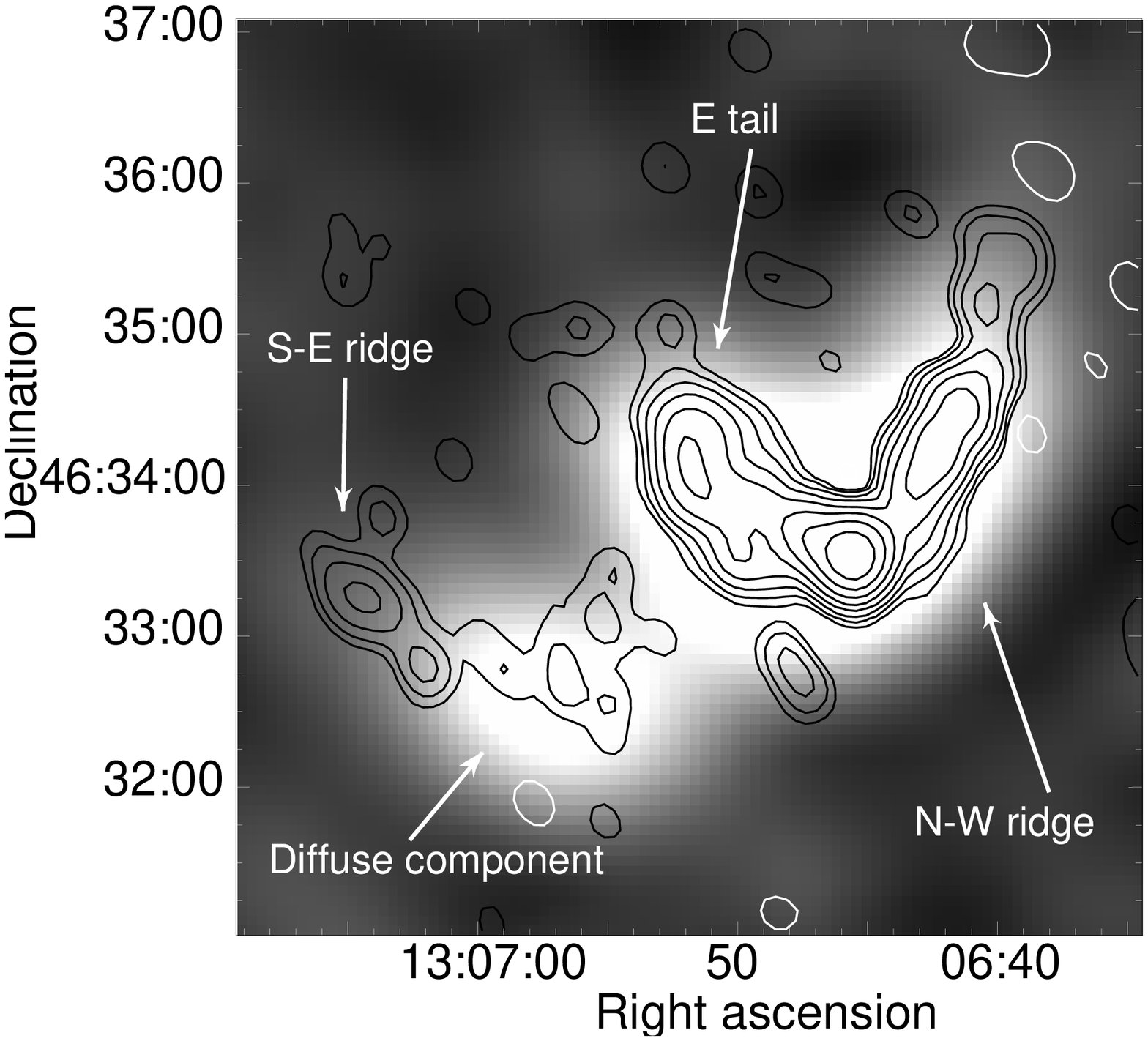}
\caption{{\it Left}: GMRT 240 MHz image of A\,1682 at the resolution of 
12.5$^{\prime\prime}\times9.2^{\prime\prime}$, p.a. 55.7$^{\circ}$.
The 1$\sigma$ rms in the image is 0.6 mJy beam$^{-1}$. Red
contours start at 2 mJy beam$^{-1}$ and scale by a factor of 2. 
Black dashed contours correspond to the $-2$ mJy beam$^{-1}$ level. Individual 
radio galaxies are labelled S1, S2 and S3. The optical image is the red plate
of the Digitized Sky Survey (DSS--2).
{\it Right}: GMRT 240 MHz contours overlaid on the VLSS 74 MHz
emission (grey scale). The resolution of the 240 MHz image is 
18.3$^{\prime\prime}\times14.0^{\prime\prime}$, p.a. 21.7$^{\circ}$.
Black contours start at 2 mJy beam$^{-1}$ ($1\sigma=0.6$ mJy beam$^{-1}$) and 
are spaced by a factor of 2. White contours correspond to the 
$-2$ mJy beam$^{-1}$ level.}
\label{fig:a1682_tot}
\end{figure*}

Our 240 MHz images are reported in Fig. \ref{fig:a1682_tot}. The left panel
shows the full resolution contours overlaid on the optical emission (red
plate of the Digitized Sky Survey DSS--2), 
while in the right panel a tapered image is overlaid on the 74 MHz 
emission visible on the VLA Low--Frequency Sky Survey (VLSS).
The N--W ridge extends $\sim 2^{\prime}$, i.e. $\sim$ 430 kpc, while 
E--tail is $\sim 380$ kpc. The S--E ridge is 
$\sim 1^{\prime}$, i.e. $\sim$ 220 kpc.
The brightness distribution of both the N--W and S--E ridge is centrally 
peaked all over their extent.
\\
Fig. \ref{fig:a1682_tot} clearly shows the presence of a new component beyond 
the features already visible at 610 MHz, labelled as ``diffuse component'' 
(right panel), and coincident with a similar feature on the VLSS. This 
component is a prominent feature also in GMRT 150 MHz images (Macario et al.
\cite{macario12}) and in LOFAR (LOw Frequency ARray) observations 
(Macario et al. in prep.).

From the low resolution image in the right panel of Fig. \ref{fig:a1682_tot}
we measured the 240 MHz flux density of the various components, and 
obtained: S$_{\rm (N-W~ridge)}=468\pm23$ mJy; S$_{\rm (E-tail)}=1126\pm56$ 
mJy (this value includes both the brightest part of the emission and the 
long tail, while the contribution of S2, associated with a 18.0 magnitude
cluster galaxy at z=0.233, Hao et al. \cite{hao10},
has been removed); 
S$_{\rm (S-E~ridge)}=63\pm4$ mJy; S$_{\rm (Diff~Comp)}=46\pm4$ mJy.
 
\begin{figure*}[htbp]
\centering
\includegraphics[angle=0,scale=0.8]{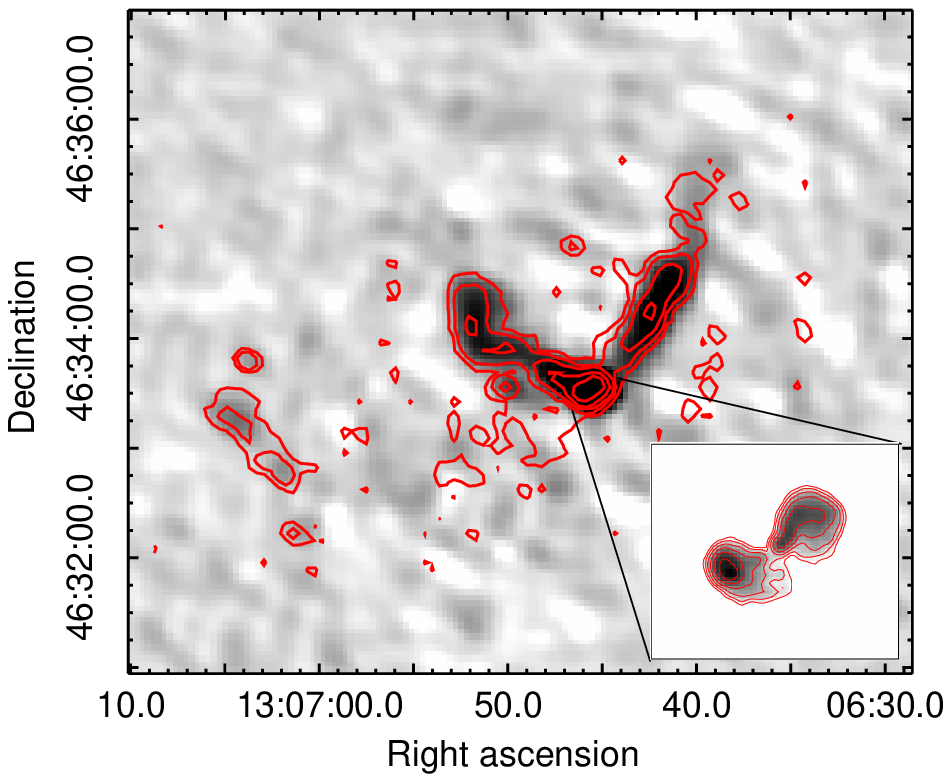}
\hspace{0.5cm}
\includegraphics[angle=0,scale=0.5]{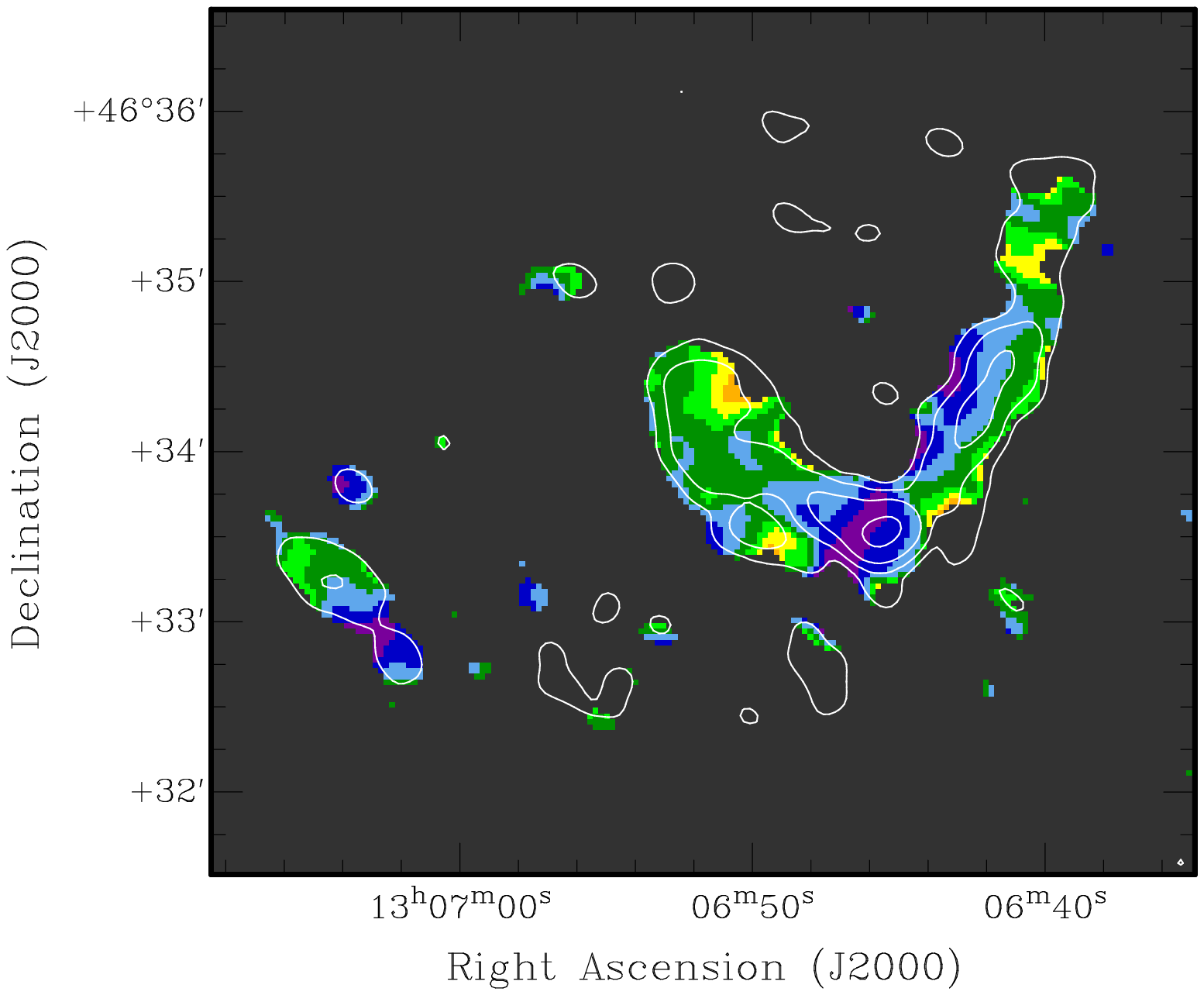}
\caption{{\it Left}: GMRT 240 MHz  grey scale image of A\,1682 with 610 
MHz contours overlaid. The 240 MHz image is the same as Fig. 
\ref{fig:a1682_tot}, right panel. The resolution of the 610 MHz image is 
6.2$^{\prime\prime}\times4.1^{\prime\prime}$, p.a. 61.2$^{\circ}$,
the 1$\sigma$ rms in the image is $25~\mu$Jy~beam$^{-1}$, the first
contour is 0.1 mJy~beam$^{-1}$, and contours are spaced by a factor of 4.
The inset in the bottom right corner is the 1.4 GHz VLA--A image of the
central BGC at the resolution of 1.1$^{\prime\prime}\times1.1^{\prime\prime}$.
Contours are spaced by a factor of 2 starting from 0.15 mJy~beam$^{-1}$.
{\it Right}: Spectral index image of A\,1682, with GMRT 240 MHz contours
overlaid (resolution 12$^{\prime\prime}\times10^{\prime\prime}$, p.a. 60$^{\circ}$)
The first contour is 3 mJy beam$^{-1}$, contours are spaced by a factor of 4.
The spectral index scale ranges from light green 
($\alpha_{\rm 240~MHz}^{\rm 610~MHZ}\sim2.3$) to violet
($\alpha_{\rm 240~MHz}^{\rm 610~MHZ}\sim0.5$).}
\label{fig:a1682_spix}
\end{figure*}

The left panel of Fig. \ref{fig:a1682_spix} shows an overlay of 
the radio emission of A\,1682 at 610 MHz (V08) and 240 MHz. The shape, 
size and brightness  distribution of the E--tail, the N--W and the S--E 
ridges are in very good  agreement at the two frequencies.

In order to throw some light on the nature of the various components of
the diffuse radio emission in this cluster, we performed a spectral
study between 610 MHz and 240 MHz.
The total spectral index of the N--W ridge, E--tail and S--E ridge between  
610 MHz and 240 MHz, derived using images with resolution  
$\sim 14^{\prime\prime}$ 
and computed integrating over the same area, are respectively:
$\alpha_{\rm 240~MHz}^{\rm 610~MHz}~{\rm (N-W~rigde)}=1.62\pm0.09$,
$\alpha_{\rm 240~MHz}^{\rm 610~MHz}~{\rm (E-tail)}=0.99^{+0.09}_{-0.10}$,
$\alpha_{\rm 240~MHz}^{\rm 610~MHz}~{\rm (S-E~rigde)}=1.53^{+0.11}_{-0.13}$.
The spectrum of E--tail actually consists of two distinct components:  
the brightest part of the emission (before the radio emission bends) has a
``normal'' spectrum, i.e. $\alpha=0.78$, typical of radio galaxies, while
the spectrum in the long tail is very steep, with $\alpha=2.1$.
\\
We point out that the 610 MHz observations we are using for this analysis 
do not belong to the same set of observations of the GMRT Radio Halo 
survey, but they are part of a longer exposure carried out two years after 
the survey observations. While it is still possible that the 610 MHz flux 
density measurements are underestimated (see Macario et al. \cite{macario10}),
a check on the S--E ridge does not suggest any missing flux at 610 MHz.
The spectral index for this feature between 240 MHz and 610 MHz is
consistent with that reported in V08 between 610 MHz and 1.4 GHz (NVSS).

To check for steepness gradients in the N--W ridge and in the E--tail
we matched the u--v coverage of the 610 MHz and 240 MHz datasets and
produced a spectral index image of the central cluster region at the 
resolution of $12^{\prime\prime}\times10^{\prime\prime}$.  Our results are 
given in the right panel of Fig. \ref{fig:a1682_spix}. The distribution of 
$\alpha_{\rm 240~MHz}^{\rm 610~MHz}$ in the N--W ridge shows a gradient parallel 
to the source major axis, from $\sim 1$ (dark blue) to $\sim 2.3$ 
(light green) going from East to West.
In the E--tail two distinct regions are clearly visible: (1) the spectral 
index shows a gradient at the South--Western
end of the structure, with $\alpha_{\rm 240~MHz}^{\rm 610~MHz}$ steepening 
from  $\sim 0.5$ at the location of the dominant double radio galaxy, 
to $\sim 1.2$ where the feature bends; (2) after the bending the spectral 
index is in the range 1.8--2.2 with a patchy distribution.

To investigate the possible connection between the double radio source 
associated with the dominant cluster galaxy and the E--tail, 
we re--analysed two short archival VLA observations at 1.4 GHz with the array 
in the A and D configuration (projects AM699 and AM469 respectively).
The VLA--A array image at the resolution of 1$^{\prime\prime}$ is 
reported in the inset in the left panel of Fig. \ref{fig:a1682_spix}.
It is clear that the BCG is a compact double, whose lobes bend on the S--W 
direction on an angular scale of $\sim 5^{\prime\prime}$. Such bending seems 
unrelated with the E--tail. The spectral index distribution along the 
E--tail seems to support this idea: no steepening is found moving away from
the nuclear emission, as it would be expected for a tailed radio galaxy.

The diffuse component south--east of the cluster centre is a new intriguing 
feature. It has the steepest spectrum. Comparison of the 240 MHz and 74 MHz 
flux density values is difficult, since the VLSS image provides very different 
values depending on the algorithm used to integrate over that portion of the 
image.
Considering the flux density measured at 610 MHz over the same region,
we obtain a spectral index $\alpha_{\rm 610~MHz}^{\rm 240~MHz}=2.09\pm0.15$. 
Finally, in V08 we reported on the presence of residual flux density in a 
cluster region of $\sim$ 1 Mpc within the centre at 610 MHz. We find a similar
situation also at 240 MHz. In Venturi et al. (\cite{venturi11b}) we carried
out an analysis of this residual emission after subtraction of the individual 
sources from the u--v data, following a procedure similar to that
adopted for A\,781 (Venturi et al. \cite{venturi11a}).
Excess flux density is confirmed at both frequencies and at 1.4 GHz.

To summarize, the radio emission at the centre of A\,1682 is very complex,
and the nature of the various features remains enigmatic.
In particular:

\begin{itemize}

\item[{\it (a)}] The size,
morphology and spectrum of the N--W ridge are consistent with our knowledge
of relic sources, however the spectral gradient between 610 MHz and 240 MHz
would suggest some merger activity from West to East, which is
not confirmed by the current X--ray data (see Fig. 7 in V08).

\item[{\it (b)}] On the basis of the high resolution VLA--A image of the BCG 
and on the spectral index distribution (Fig. \ref{fig:a1682_spix}), it seems 
unlikely that the E--tail is associated with the BCG, but at present there 
are no alternative suggestions for its origin. We note that the small scale
radio emission of the BCG hints at a motion in the Southwest--Northeast 
direction. 

\item[{\it (c)}] The size of the S--E ridge is only $\sim$ 220 kpc, and it
is considerably smaller than the relics known so far.

\item[{\it (d)}] The nature of the diffuse component is unclear. One 
possibility is that it is the brightest part of an underlying steep spectrum 
radio halo. Alternatively, it could be a dying radio galaxy.
\end{itemize}

A detailed multifrequency study of this cluster,
based on deep observations at 74 MHz (VLA--A, Venturi et al. 
\cite{venturi11b}), at 150 MHz (GMRT, Macario et al. \cite{macario12}) 
and 325 MHz (GMRT), will allow to us classify the diffuse extended sources 
and throw a light on their nature (Dallacasa et al. in preparation).

\subsection{Z\,2661}

\begin{figure*}[htbp]
\centering
\includegraphics[angle=0,scale=0.95]{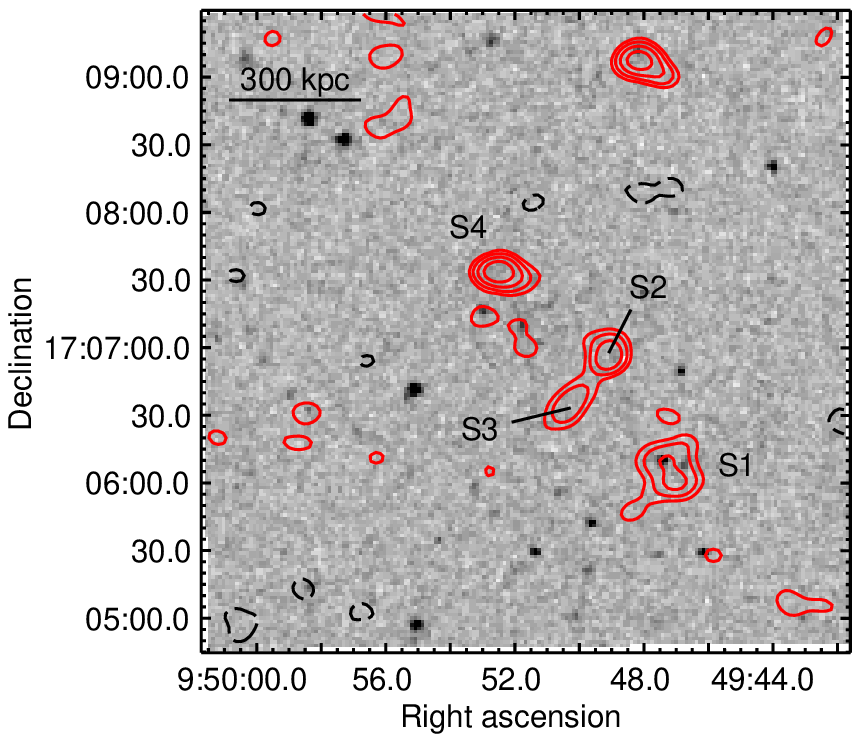}
\hspace{0.2cm}
\includegraphics[angle=0,scale=0.97]{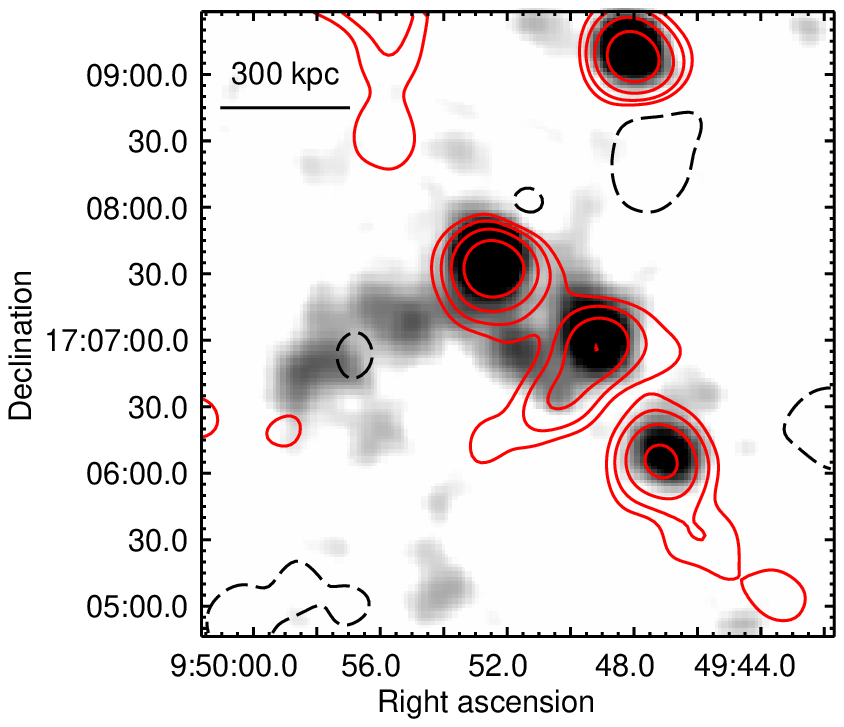}
\caption{{\it Left}: GMRT 325 MHz radio contours of the emission from Z\,2661
at the full resolution (10.8$^{\prime\prime}\times8.4^{\prime\prime}$,
p.a. $-83^{\circ}$), overlaid on the POSS--2 red image. The 1$\sigma$
noise level is 0.12 mJy beam$^{-1}$. Red contours start at 0.36 mJy beam$^{-1}$ 
and then scale by a factor of 2. Black dashed contours 
correspond to the $-3\sigma$ level. Individual radio galaxies are
labelled from S1 to S4. {\it Right}: 325 MHz contours at the
resolution of 25.0$^{\prime\prime}\times25.0^{\prime\prime}$,
p.a. $0^{\circ}$, overlaid on the GMRT 610 MHz image 
(17.7$^{\prime\prime}\times17.4^{\prime\prime}$) from V08.
Red contours start at $+3\sigma=$0.6 mJy beam$^{-1}$ and then scale by a 
factor 2. The $-3\sigma$ level is shown as black dashed contours. 
The rms noise level in the 610 MHz image is 65 $\mu$Jy beam$^{-1}$.}
\label{fig:z2661}
\end{figure*}

Z\,2661 is the  second most luminous cluster in the GMRT Radio Halo Cluster 
sample, with L$_{\rm X, [0.1-2.4 KeV]}=1.78\times10^{45}$ erg s$^{-1}$. 
V08 reported the
presence of very faint candidate diffuse emission at the cluster centre 
(S$_{\rm 610~MHz}\sim$5.9 mJy). Our 325 MHz observations failed to
detect diffuse emission. Fig. \ref{fig:z2661} shows the central
cluster region at two different resolutions. Only the radio emission
from the individual sources is clearly detected, and no significant
excess flux density is measured over the same region covered by the
diffuse emission at 610 MHz. 


\subsection{The remaining clusters in the sample}

In this section, for completeness and to help the discussion, we briefly 
summarize the most important radio properties of the four clusters followed 
up with the GMRT at 325 MHz and 240 MHz (Table 1), already published in 
separate papers.

\subsubsection{A\,521: the prototype ultra-steep spectrum radio halo}

The low frequency observations of A\,521, carried out at 240 MHz and
325 MHz to study the cluster
relic (Giacintucci et al. \cite{giacintucci08}), led to the discovery of  
a very steep spectrum 
giant radio halo at the cluster centre (Brunetti et al. \cite{brunetti08}). 
The source was detected in very deep 1.4 GHz VLA observations 
(Dallacasa et al. \cite{dallacasa09}), which provided
$\alpha^{\rm 1.4~GHz}_{\rm 240~MHz}=1.86\pm0.08$, and further imaged with the GMRT at 
150 MHz (Macario et al. \cite{macario12}).
The spectrum of the relic is a power law with 
$\alpha^{\rm 4.8~GHz}_{\rm 240~MHz}=1.48$ (Giacintucci et al. \cite{giacintucci08}).

\subsubsection{The ultra--steep spectrum radio halo in A\,697}

The GMRT 325 MHz observations of the faint radio halo in A\,697 detected
at  610~MHz (V08) revelaed that it is another example of very steep spectrum
source. The 325 MHz images and a detailed study of the radio halo and the 
properties of the hosting cluster were presented in Macario et al. 
(\cite{macario10}), who reported a flux density value 
S$_{\rm 325~MHz}=47.3\pm2.7$ mJy and a spectral index 
$\alpha_{\rm 325~MHz}^{\rm 1.4~GHz}=1.8\pm0.1$. This value is in line
within the errors with the analysis carried out in van Weeren et al.
(\cite{vanweeren11}), who reported $\alpha=1.64\pm0.06$ on the basis
of deep 1.4 GHz images.
The steep spectrum has been confirmed by the analysis of GMRT 150 MHz data   
(Macario et al. \cite{macario11b}; Macario et al. \cite{macario12}).
The linear size of the source at this frequency is $\sim$ 1.3 Mpc,
considerably larger than at higher frequencies. This giant radio halo
has a smooth surface brightness over its all extent.
Literature X--ray and optical data suggest that the cluster is unrelaxed, 
and even though the available information does not allow a detailed study, 
either a multiple merger or a major merger along the line of sight 
are the possible scenarios. 

\subsubsection{Complex radio emission in A\,781}

A\,781 is an intriguing case. Beyond a number of individual galaxies,
the central part of the cluster is dominated by a diffuse source located 
at the border of the X--ray emission, classified as a radio relic with 
spectrum $\alpha_{\rm 325~MHz}^{\rm 1.4~GHz}=1.25\pm0.06$ (V08; 
Venturi et al. \cite{venturi11a}).

On the basis of a detailed radio/X--ray analysis carried out by Cassano et 
al. (\cite{cassano10a}), the cluster was classified as ``outlier'' in the 
quantitative correlations relating cluster mergers and the presence of a 
radio halo: the cluster shows a high degree of disturbance but lacks a 
radio halo. Our GMRT 325 MHz follow--up observations revealed 
the presence  of residual emission at the cluster centre, after subtraction 
of the  individual sources (Venturi et al. \cite{venturi11a}). Such excess 
flux density sums up to $\sim$ 20 mJy and is visible at the 2$\sigma$ level. 
Our findings did not confirm the 1.4 GHz detection of the radio halo claimed 
in Govoni 
et al. (\cite{govoni11}), unless the source has an unplausibly flat spectrum
($\alpha_{\rm 325~MHz}^{\rm 1.4~GHz}\le0.5$, see Venturi et al. 
\cite{venturi11a} for a complete discussion).

An attractive possibility is that the residual emission reveals the presence 
of an underlying very low surface brightness radio halo. Allowing for all 
the uncertainties in our analysis, an accurate comparison between the 325 
MHz and 610 MHz images suggests that the 325 MHz residuals could be 
consistent with a diffuse source with spectrum steeper than $\alpha >$ 1.5.

\subsubsection{RXCJ\,2003.5$-$2323}

The cluster is one of the most distant in our sample (z=0.317) and hosts
a giant radio halo (LLS$\sim$1.4 Mpc), discovered with the
GMRT at 610 MHz (V07). The cluster been extensively followed--up by us in 
the radio, X--ray and optical band (Giacintucci et al. \cite{giacintucci09}). 
\\
The radio halo is very powerful, and its spectrum can be well fitted by 
a single power law with spectral index 
$\alpha_{\rm 240~MHz}^{\rm 1.4 GHz}=1.27^{+0.18}_{-0.08}$.
It has a regular shape and its size is very similar from 240 MHz to 1.4 GHz. 
Its brightness distribution is characterised by clumps and filaments. Our 
multiwavelength study supports the scenario of a merger--driven formation 
for this giant radio halo.


\begin{figure*}[htbp]
\includegraphics[angle=0,scale=0.5]{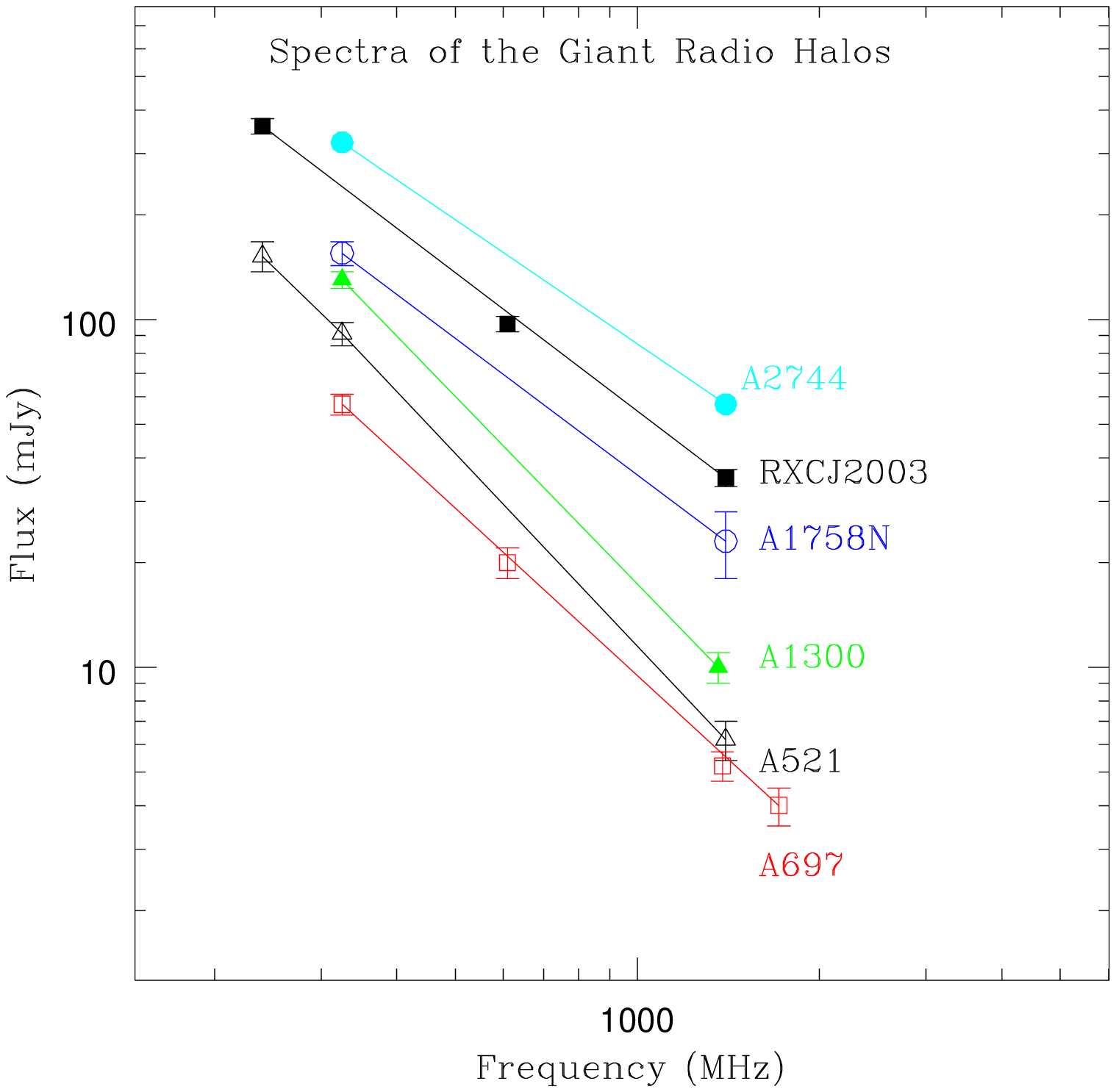}
\hspace{0.5cm}
\includegraphics[angle=0,scale=0.5]{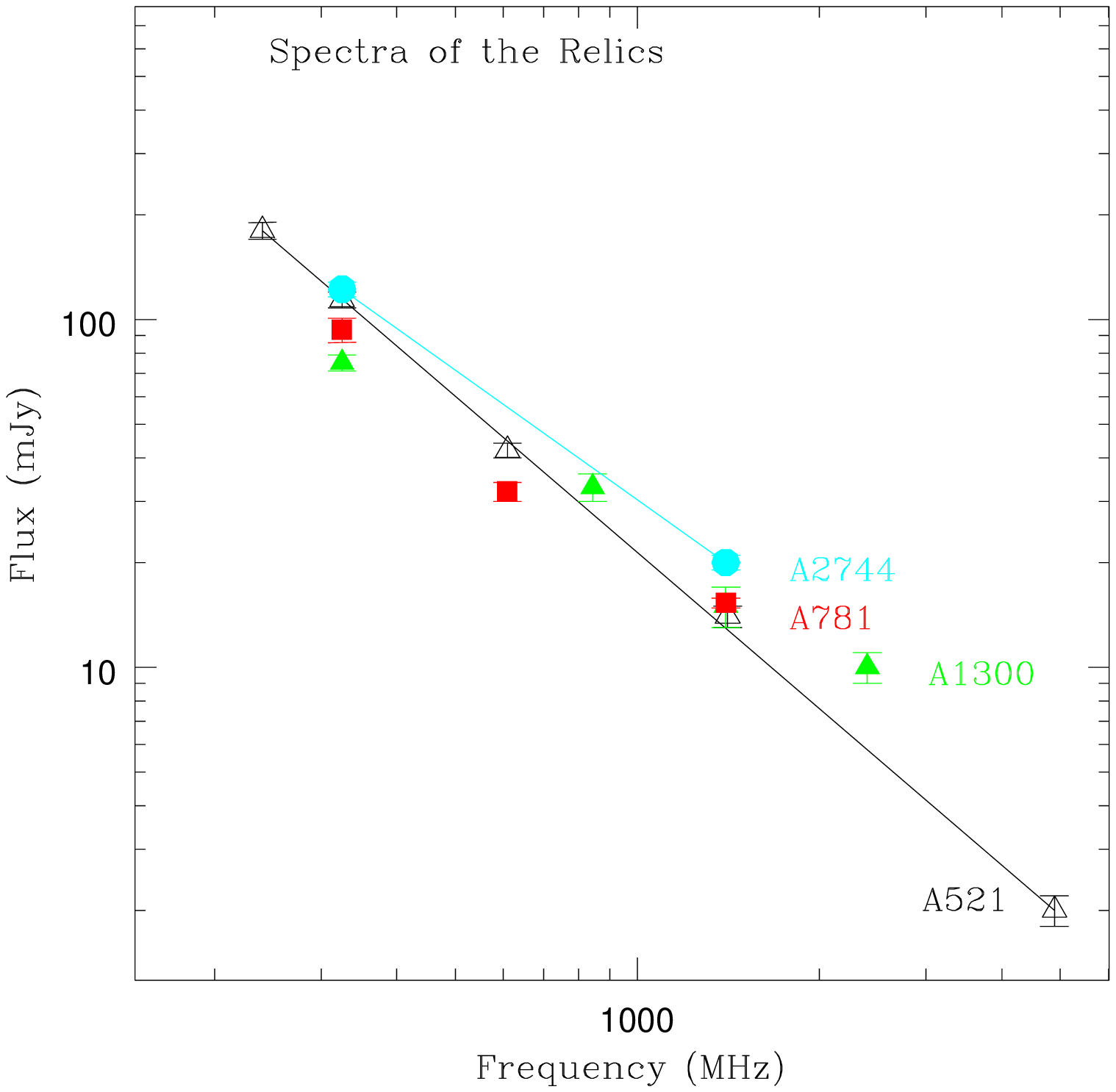}
\caption{{\it Left}: Spectra of the giant radio halos in the GMRT
Radio Halo Survey. The 325 MHz data points are reported in Table 4. The
remaining data are taken from: A\,697 Macario et al. (\cite{macario10}) 
and van Weeren et al. (\cite{vanweeren11}); A\,521 Brunetti et al. 
(\cite{brunetti08}); RXCJ2003.5--2323 Giacintucci et al.
(\cite{giacintucci09}).
For A\,1300 see Sect. 3.2.1; for A\,1758N see Sect. 3.1.2; for A\,2744
see Sect. 3.2.2.  
{\it Right}: Spectra of relics from the GMRT Radio Halo Survey.
The 325 MHz data points are reported in Table 4. The remaining data  are 
taken from: A\,521 G08; A\,781 Venturi et al. (\cite{venturi11a}); A\,1300
R99; A\,2744 see Sect. 3.2.2.
In both panels the lines are drawn to help the visual inspection and
are not the best fit power law.}
\label{fig:spectra}
\end{figure*}


\section{Discussion}\label{sec:disc}

The main goal of our GMRT 325 MHz follow--up program of the halos, 
relics and candidates in the GMRT cluster sample (i.e. redshift range 
z=0.2--0.4, X--ray luminosity 
L$_{\rm X}{\rm 0.1-2.4~keV}>5\times10^{44}{\rm erg s^{-1}}$, 
$\delta>-30^{\circ}$, V07 and V08) was to perform high sensitivity imaging
of diffuse cluster sources at a key frequency for our understanding of
their origin. Both halos and relics are easily detected at 325 MHz, 
due to their steep spectrum and increased brightness with respect to the 
GHz observing frequencies. Compared to the few other interferometers 
operating at frequencies of the order of few hundred MHz, the GMRT has 
the advantage to allow accurate subtraction of the individual sources 
usually embedded in the diffuse emission, thus allowing a much more 
precise flux density measurement of the diffuse cluster sources, 
as well as much more reliable imaging of their fine scale features. This 
is critical to perform a detailed comparison with the distribution of 
the X--ray emission, and to derive reliable integrated spectra.
As a matter of fact, integrated spectra of radio halos still suffer from
many uncertainties, mainly due to the miscellanoeus collection of data, 
i.e. different arrays, resolutions, and procedures in the subtraction
of individual embedded sources (see Venturi \cite{venturi11}).

Observations at frequencies $\nu~\le~325$ MHz clearly show that the diffuse
radio emission in galaxy clusters is very complex: new features 
show up, beyond the canonical classification of radio halos and relics,
and the appearance of radio halos may change with frequency, both in some
morphological details, and in the overall size. Some of the most 
relevant findings in our work are summarized and discussed in the following
sections.

\subsection{Morphology of radio halos}

\subsubsection{Size versus frequency}

Our results clearly show that our view of radio halos changes  
at low frequencies.
In some cases, the size of radio halos is very similar going from 1.4 GHz 
down to 325 MHz and 240 MHz. A clear example is 
RXCJ\,2003.5--2323 (Giacintucci et al. \cite{giacintucci09}), and to a less
extent A\,2744. The total size of A\,1300 is 
similar in our image and those in R99, even though the different angular 
resolution and sensitivity do not allow an accurate comparison.\\
On the other hand, some radio halos with spectral index steeper than
$\alpha_{\rm 325~MHz}^{\rm 1.4~GHz}>1.6$ (see also next section) are more extended at 
frequencies of few hundred MHz, being barely detected at 1.4 GHz, as is
the case for A\,521 (Dallacasa et al. \cite{dallacasa09}).
\\
While it is tempting to suggest that the ``GHz halos'', i.e. those with 
``normal'' ($\alpha \sim 1.2\div1.3$) spectrum, tend to have size and shape 
similar at all frequencies, and the ``low frequency halos'', i.e. those with 
very steep spectrum  ($\alpha \ge1.6$) show increasing size at decreasing 
frequency, as is the case for A\,697 (see Macario et al. \cite{macario10} and
\cite{macario11b}, and van Weeren et al. \cite{vanweeren11}), it is possible 
that this difference 
is the result of the limited sensitivity of the current interferometers.
The improved sensitivity of the EVLA at GHz frequencies and of the LOFAR
below 240 MHz will allow to carefully address this point, and explore the 
possibility that the dependence of the size with frequency is the result of 
the radial spectral steepening.

\subsubsection{The brightness distribution and radio/X comparison}

An important outcome of our 325 MHz survey concerns the brightness 
distribution of radio halos, which differs from case to case: 
A\,697 is centrally peaked;
A\,2744 is asymmetrically peaked; most halos have a rather ``clumpy''
brightness distribution (A\,521, Brunetti et al. \cite{brunetti08}; A\,1300; 
A\,1758N;  RXCJ\,2003.5--2323, Giacintucci et al. \cite{giacintucci09}).
Other known clusters in the literature show similar properties: the Coma 
cluster is centrally peaked (see for instance the 325 MHz image reported in 
Brown \& Rudnick \cite{shea11}), while A\,754 is clumpy (Macario et al. 
\cite{macario11a}).

The comparison of the radio and X--ray brightness distribution provides
further insightful information. ``Clumpy'' radio halos are usually found in 
clusters with inhomogneous X--ray distribution, as is the case for A\,521, 
A\,1300 (see right panel of Fig. 3) and  RXCJ\,2003.5--2323. 
On the other hand, the peak in the radio halo brightness distribution
may coincide with the thermal peak, as is the case for A\,697 (Macario et 
al. \cite{macario10}), or be misplaced from it, as it is clear in A\,1758N 
and A\,2744 (right panel of Figs. 2 and 4 respectively).

These observational differences may contain important information of the 
type of merger and on the interplay between the cluster dynamical activity 
and the non--thermal emission. This result should be taken into account when 
addressing the origin of radio halos.

\subsection{Integrated spectra of halos and relics}

One of the most outstanding results of our low frequency follow--up of  the 
GMRT Radio Halo survey has been the discovery of  ultra--steep spectrum radio 
halos, i.e. A\,521 and A\,697 (see Sections 3.5.1 and 3.5.2 respectively).
To highlight the different spectral slope of the spectrum of these two clusters
compared to the others surveyed with the GMRT at 325 MHz, in the left panel 
of Fig. \ref{fig:spectra} we report the spectra of the giant radio halos in 
A\,2744, A\,521, A\,697, A\,1300, A\,1758N, RXCJ\,2003.5--2323.
It is worth pointing out that A\,521, A\,697 and RXCJ\,2003.5--2323 have been 
analysed using the same approach on the datasets at all frequencies. 
This is critical if we 
consider all the uncertainties in the subtraction of the contribution
of individual sources embedded in the diffuse radio emission.
The lines are not the best fit, and are drawn only to help the 
visual inspection.

Even considering only the clusters in the GMRT sample, it is clear that
the observed spectral index of radio halos is quite different from case to
case, and it goes from $\alpha_{\rm 325~MHz}^{\rm 1.4~GHz}\sim1.2$ (A\,2744) to 
$\sim$~1.9 (A\,521), with a spread of at least $\Delta\alpha\sim$0.7. 
This reflects into a spread in the energy distribution of the relativistic 
particle population
of the order $\Delta\delta^{\rm e^{\pm}}$~$\sim$~1.4, which needs to be explained. 
One possibility is that a variety of mechanisms produces relativistic particles
with different energy distributions. 
More likely, the broad range of observed synchrotron spectra suggests that 
the spectrum of the relativistic particles is not a power law, and that it
is the result of the combined effect of a high energy break in the electron 
energy distribution and an inhomogeneous magnetic field over the cluster volume.
A break in the spectrum of the emitting particles is predicted by turbulent 
re--acceleration models, or more generally, when the time scale of particle
re-acceleration is comparable to that of radiative losses 
(10$^7$--10$^8$ years).
\\
\\
The observational properties of the relics in the GMRT radio halo survey 
(A\,521, A\,781, A\,1300, A\,2744) are more homogeneous.
In all cases their morphology, brightness distribution and extent are fairly 
consistent at all frequencies. The only exception is A\,781, which is 
more extended at 325 MHz, compared to 610 MHz and 1.4 GHz (Venturi et al.
\cite{venturi11a}). The spectral index $\alpha_{\rm 325~MHz}^{\rm 1.4~GHz}$
is in the range 1.25 (A\,781) -- 1.48 (A\,521).\\
We do not include the double relic cluster RXCJ\,1314.4--2515 in these
spectral considerations, 
due to the poor quality of the 325 MHz images. This cluster has been
reobserved with the GMRT at 325 MHz and a separate paper is in progress
(Giacintucci et al. in prep). The spectral index between 610 MHz and 1.4 GHz
is $\alpha_{\rm 610~MHz}^{\rm 1.4~GHz}=1.40\pm0.09$ and $ 1.41\pm0.09$ for the
western and eastern relic respectively (V07).
\\
In the right panel of Fig. \ref{fig:spectra} we report the spectra of the 
relics in A\,2744, A\,521, A\,781, A\,1300. 
The spread in flux density for these sources is much narrower compared 
to halos. The lines are drawn to help the visual inspection. Due to the 
small number of available relic radio spectra we cannot conclude 
that this is a general behaviour.

\subsection{Diffuse cluster emission beyond halos and relics}

A number of diffuse sources, which do not fit either into the radio halo 
or relic class, have been detected in some clusters. Our findings are 
summarized here below. 

\begin{itemize}

\item[--]
A clear ``bridge'' of emission connects the radio halo and the relic in 
A\,2744 (right panel of  Fig. \ref{fig:a2744_tot}) and in RXCJ\,1314.4--2515 
(V07); a similar feature is visible in A\,521 (Brunetti et al. 
\cite{brunetti08}) and A\,1300 (right panel of Fig. \ref{fig:a1300_tot}) in the
region between the halo and the relic.
Radio bridges have been so far observed only in very few famous cases, such as 
the Coma cluster (Kim et al. \cite{kim89}; Brown \& Rudnick \cite{shea11})
and A\,3667 (Carretti et al. \cite{carretti12}).
Projection effects cannot be ruled out, however an attractive explanation
for these features assumes
that they may origin from merger shock re--accelerated electrons, which move 
downstream as the shock advances through the ICM and are further accelerated
by the turbulence injected by the merger in the ICM closer to the cluster 
centre (Markevitch \cite{markevitch10}).

\item[--]
Extended emission, which cannot be easily classified, was detected 
in A\,1300 (the ``candidate relic'', see Sect. 3.2.1). This source is 
arc--shaped and has a projected linear size of the order of $\sim~$700 kpc 
and radio power logP$_{\rm 325~MHz}=6.9\times10^{24}~{\rm W Hz^{-1}}$ (assuming 
that it is at the cluster distance), and it is located at the periphery of 
the X--ray emission.

\item[--]
A\,1682 is the most enigmatic cluster in our study. Three filamentary features 
with no obvious optical counterpart
are found: the N--W ridge, the E--tail and the S--E ridge. The latter is 
located at the boundary of the X--ray emission, but it is much smaller 
(at least in projection) than the typical size reported for relics, i.e. 
only $\sim~$220 kpc, and has a radio power 
logP$_{\rm 240~MHz}=9.3\times10^{24}~{\rm W Hz^{-1}}$ if located at the distance
of A\,1682.
The morphology, size and radio power of the N--W ridge argue in favour
of a possible relic, but the direction of the steepening in the spectral index
distribution is a puzzle.
Our analysis of the E--tail disfavours the idea that it is related to the 
central double radio galaxy, even though at present there are no alternative 
explanations for its nature. Finally, diffuse very 
low surface brightness emission is visible south of the cluster centre, with 
total radio power logP$_{\rm 240~MHz}=6.8\times10^{24}~{\rm W Hz^{-1}}$ (assuming
it is located at the distance of the hosting cluster). This radio emission
could be the brightest part of an underlying more extended very low surface 
brightness radio halo.

\item[--]
Our data  suggest the presence of diffuse emission in A\,781 just below the 
sensitivity threshold of the observations (Venturi et al. \cite{venturi11a}, 
see Table 5). This cluster deserves further investigation. 

\end{itemize}
All the above results reinforce the overall finding that the diffuse cluster
radio emission is more complex than the usual ``radio halo'' and ``relic''
classification when observed at frequencies of the order of $\le$ 325~MHz.

\begin{figure*}[htbp]
\centering
\includegraphics[angle=0,scale=0.4]{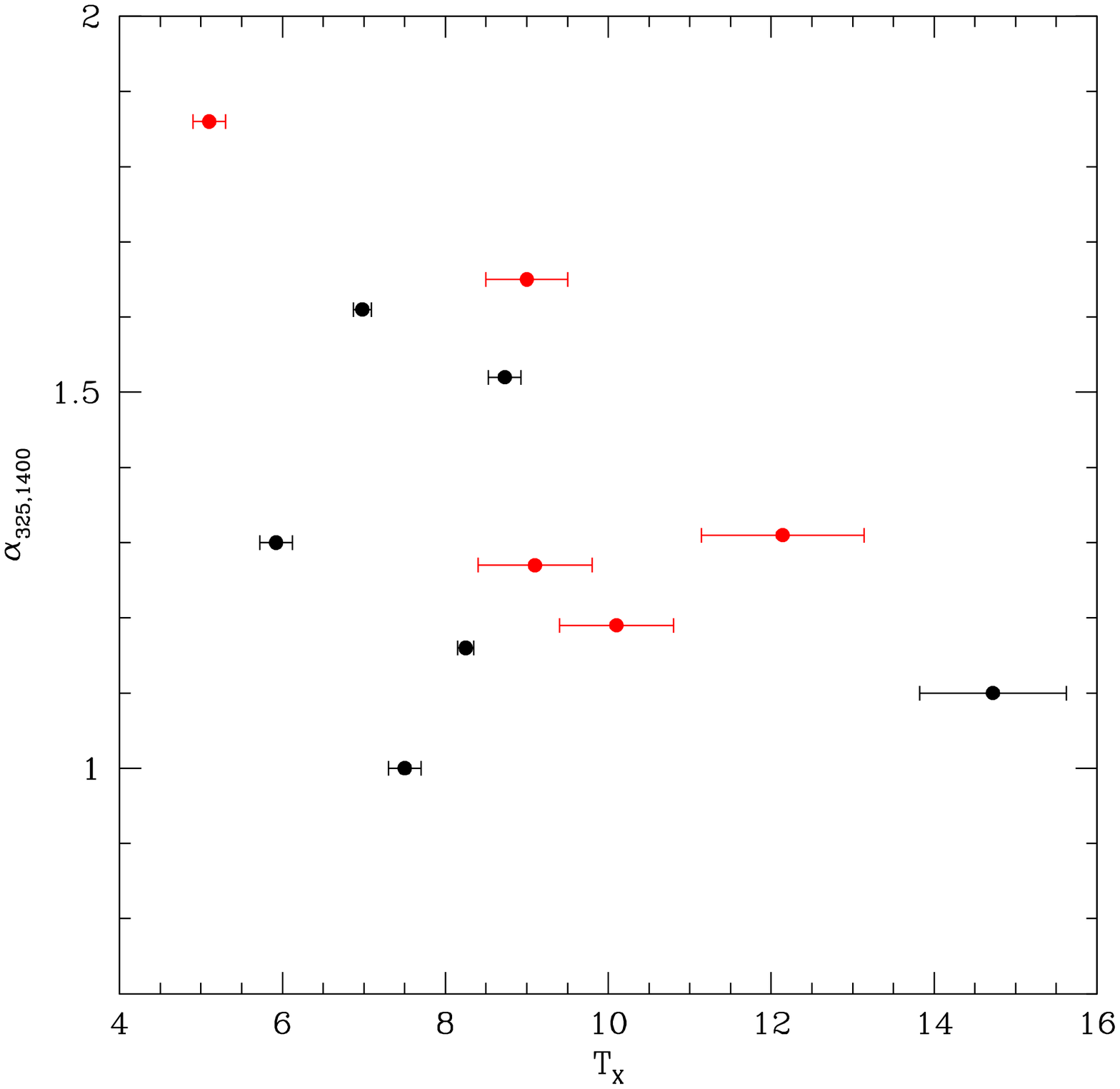}
\hspace{0.5cm}
\includegraphics[angle=0,scale=0.4]{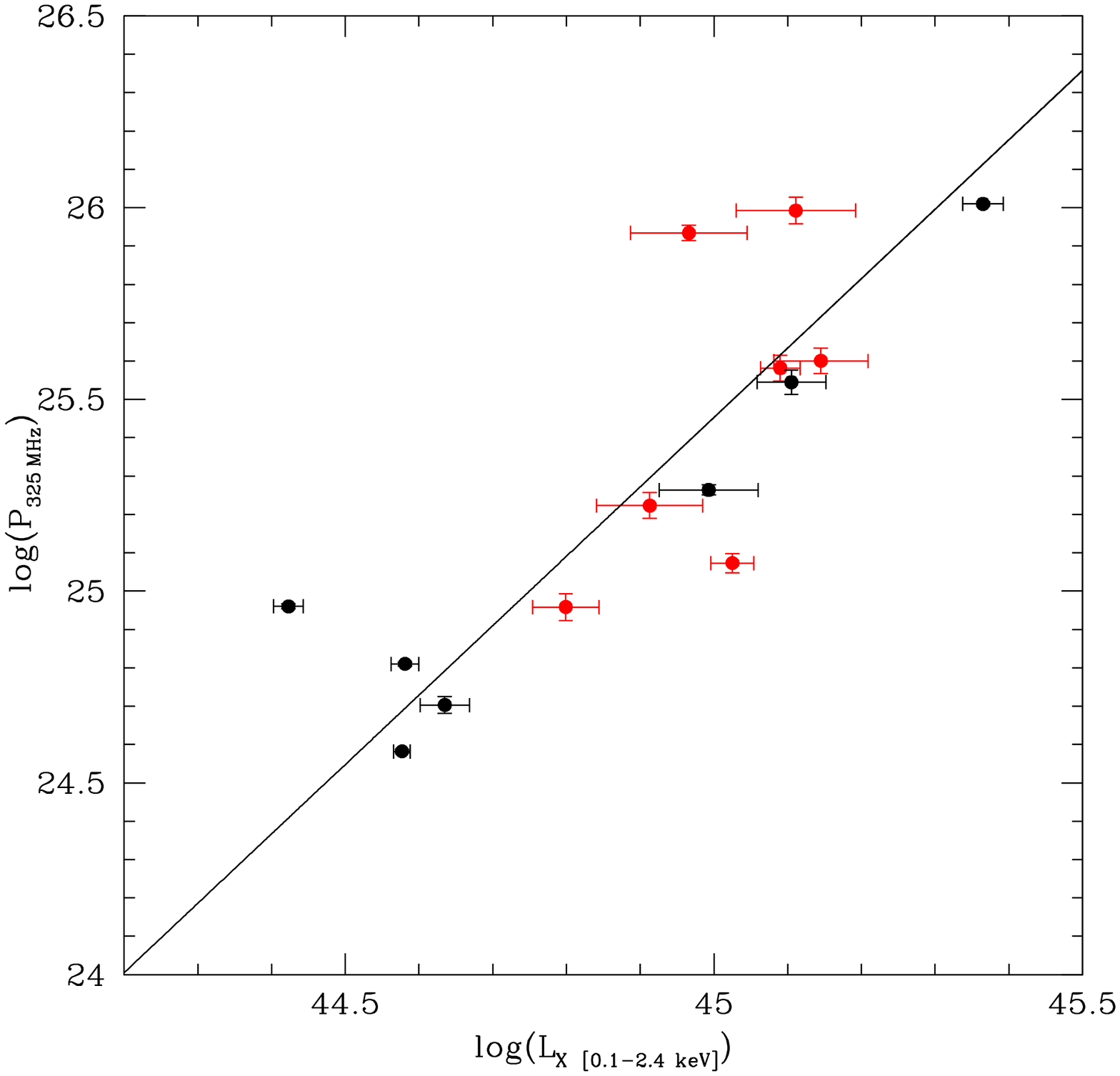}
\caption{{\it Left}: Spectral index of the radio halos presented in this paper
and collected from the literature in the range 325 MHz--1.4 GHz plotted
as function of the cluster X--ray temperature.
{\it Right}: LogL$_{\rm X}$--LogP$_{\rm 325~MHZ}$ correlation. The red points in
each plot are the clusters from the GMRT 325 MHz follow-up.}
\label{fig:correlations}
\end{figure*}

\subsection{Statistical correlations for radio halos at low frequency}

The low frequency follow--up of the GMRT Radio Halo Cluster Survey 
has doubled the number of galaxy clusters with high sensitivity
imaging of the radio halo at 325 MHz, bringing them to 14. We are aware
that the statistics is still limited, but the numbers are significant 
enough to start investigating some correlations.


\subsubsection{Spectral index of radio halos and cluster temperature}

Feretti et al. (\cite{feretti04}), and more recently Giovannini et al.
(\cite{giovannini09}), claimed a possible 
correlation between the cluster temperature and the radio halo spectral 
index, suggesting that hotter clusters host radio halos with flatter 
spectrum. If confirmed, this trend would provide important constraints
on the origin of the emitting particles. 

To investigate this point, to our new results we added all the remaining 
clusters in the literature with spectral information between 325 MHz and 
1.4 GHz (A\,2163, Coma, A\,2255, A\,665, A\,2256 and A\,754). Our 
analysis does not include A\,209 and A\,1300 (see Sect. 3.1.1 and 3.2.1 
repsectively).
Note that we used the updated values of 
$\alpha_{\rm 325~MHz}^{1.4~GHz}=1.65$ for A\,697 (van Weeren 
et al. \cite{vanweeren11}) and $\alpha_{\rm 325~MHz}^{1.4~GHz}=1.3$ for A\,2255
(Pizzo \& de Bruyn \cite{pizzo09}). 
These two values are considerably different than those used in the analysis
of Giovannini et al. (\cite{giovannini09}), i.e. 1.2 and 1.7 respectively
for A\,679 and A\,2255.
\\
In the left panel of Fig. \ref{fig:correlations} we report the distribution 
of the clusters in the  $\alpha_{\rm 325,1400}$--T plane.
We do not find a clear correlation between these two quantities. 
Nevertheless, at a preliminary level our plot indicates that:
\begin{itemize}

\item[(1)] clusters with T$<$10 keV host halos both with flat ($\alpha\sim$1) 
and steep ($\alpha\sim$1.8) spectrum;

\item[(2)]
clusters with T$>$10 keV host halos with spectra flatter than $\alpha\sim$1.3.

\end{itemize}

This behaviour is qualitatively consistent with the expectations of the
re--acceleration models, which predict that radio halos have different 
spectra depending on the energy released into particle re--acceleration
during mergers. In this case, from simple enery arguments, the fraction 
of halos with ``flatter'' spectra increases with the mass of the host
cluster. In particular, if we restrict to the mass (temperature) range
of our sample, moderately massive clusters (T$\sim 5~-~8$ keV) would 
statistically generate both ``steep'' and ``flat'' halos, while massive
(T$>$ 8 keV) clusters would preferentially generate flatter halos
(see Cassano et al. \cite{cassano06} and Cassano et al. \cite{cassano10a}).
We are aware that information on more clusters is necessary
to test any possible correlation between these two quantities.

\subsubsection{LogL$_{\rm X}$--LogP$_{\rm 325~MHz}$ correlation}

Beyond the GMRT clusters presented and discussed in this paper, we collected 
all the remaining clusters in the literature with high sensitivity
imaging at 325 MHz (\,A2163, Coma, A\,2255, A\,665, A\,2256, A\,754, A\,2219)
and derived the logL$_{\rm x}$--logP$_{\rm 325~MHz}$  correlation,
which is shown in the right panel of Fig. \ref{fig:correlations}.
This analysis was presented in Kempner \& Sarazin (\cite{KS01}) using 
WENSS data, and in Rudnick \& Lemmerman (\cite{rudnick09}). For the first
time we derive this correlation  based on deep pointed observations 
with accurate subtraction of foreground and background sources projected
onto the radio halo emission.
\\
The best fit value of the slope is 1.81$\pm$0.28, consistent within the errors 
with what is found at 1.4 GHz, i.e. 2.06$\pm$0.20 
(Brunetti et al. \cite{brunetti09}).
\\
Given the small size of the sample available at 325 MHz more data are needed 
to confirm the presence of the correlation and to constrain its slope. 
Cassano (\cite{cassano10}) showed that in the framework of the turbulent 
re--acceleration scenario
the correlation becomes steeper and broader at the LOFAR frequencies and 
sensitivities.

\section{Conclusions}\label{sec:summary}

The GMRT Radio Halo Cluster Survey has been the first step towards the
investigation of the statistical properties of radio halos in galaxy
clusters. The most relevant results can be summarized as follows:
(a) the bimodal distribution of galaxy clusters with respect to the presence
of a radio halo (Brunetti et al. \cite{brunetti07}) clearly shows that 
radio halos are not ubiquitous in clusters. 
The distribution of galaxy clusters in the logL$_{\rm X}$--logP$_{\rm 1.4~GHz}$
plot provides an important piece of information for the the theoretical
models, suggesting that radio halos are transient radio sources 
(Brunetti et al. \cite{brunetti09}, En\ss{}lin et al \cite{ensslin11})
\footnote{Using a SZ--selected  sub--sample of the GMRT clusters, 
obtained from the first Planck cluster catalogue, Basu (\cite{basu12}) 
found that the cluster radio bimodality becomes considerably weaker in the 
Y$_{\rm ZS}$--logP plot, rising additional questions on the evolution
of radio halos.};
(b) first quantitative correlations
have been found between the dynamical state of clusters and the presence
of halos, or lack thereof (Cassano et al. \cite{cassano10a}). This result
confirms that mergers play the most important role in the formation of
radio halos;
(c) a relevant byproduct of the survey was the discovery of ultra--steep
spectrum radio halos. This poses new compelling constraints on the
origin of these sources.

The results of the low frequency follow--up of radio halos and relics in
the GMRT radio halo cluster sample reported in this paper, not only reinforce 
the radio halo/cluster merger connection, but also pose new important 
questions. 

The study of the integrated synchrotron spectrum of the radio halos in our
sample clearly shows a wide range of values for the spectral index, with 
$\Delta\alpha_{\rm 325~MHz}^{\rm 1.4~GHz}\sim 1.2\div1.9$, i.e.  a difference 
of $\sim$1.4 in the power of the distribution of the radiating particles. 
This result, together with our points (a) and (b) above,
is consistent with the expectations of the turbulent re--acceleration
model, which makes the unique prediction of a high frequency 
break in the spectrum of the electrons emitting at radio frequencies.
This reflects into a range of spectral slopes in the spectra of
radio halos, including the ultra steep ones.
At a more general level, these observational results suggest that 
inefficient mechanisms of particle acceleration activated during cluster
mergers play an important role in the generation of radio halos. 

We found differences in the surface brightness distribution of the 
radio halos. Some of them are clumpy, others are centrally peaked and 
in other cases the isodensity contours are asymmetrically peaked. In some
clusters we found a misplacement between the radio and X--ray peaks
of emission, which deserves further investigation.

Bridges of emission, connecting (at least in projection) the relic and
the radio halo emission were found in A\,521, A\,1300, A\,2744 and 
RXCJ\,1314.5--2515. Such features can be explained in the context
of merger shocks and turbulence during merger processes (Markevitch 
\cite{markevitch10}).

Our 325 MHz survey has doubled the number of clusters with high sensitivity
imaging of the diffuse cluster scale emission at this frequency. This
allowed us to start investigating some statistical properties of radio halos
at low frequencies. 
For the first time we derived the correlation between the radio power at 
325 MHz and the X--ray cluster luminosity by means of deep pointed
observations. The slope is consistent within the errors with that derived 
at 1.4 GHz, however the correlation needs to be constrained 
on more solid basis with a larger number of data.

Finally, we investigated a possible trend between the radio halo spectral 
index and the temperature of the hosting cluster using the new data 
presented in the paper and updated information from the literature. 
With the data available at present, we cannot confirm the existence of 
the $\alpha$--T correlation claimed in Giovannini et al. (\cite{giovannini09}).

Our study shows that systematic high sensitivity low frequency imaging of 
diffuse cluster radio sources is the next step which needs to be taken to 
further improve our  knowledge of the mechanisms at the origin of their 
formation. 

\medskip
{\it Acknowledgements.}
We thank the staff of the GMRT for their help during 
the observations. GMRT is run by the National Centre for Radio Astrophysics 
of the Tata Institute of Fundamental Research. We acknowledge financial 
contribution 
from the Italian Ministry of University and Research, from MIUR grants
PRIN2005 and 2006, from PRIN--INAF2005 and PRIN--INAF2008 and from 
contract ASI--INAF I/023/05/01. 
Support for S.G. was provided by NASA through Einstein Postdoctoral
Fellowship Award Number PF0-110071 issued by the Chandra X-ray
Observatory Center, which is operated by the Smithsonian Astrophysical 
Observatory.


\begin{thebibliography}{}

\bibitem[2003]{bacchi03}
Bacchi, M., Feretti, L., Giovannini, G., et al.,
2003, A\&A, 400, 465

\bibitem[2012]{basu12}
Basu, K., 2012, MNRAS Letters, 421L, 112

\bibitem[1999]{blasi99}
Blasi, P., Colafrancesco, S., 1999, A. Ph., 12, 1999

\bibitem[2004]{boeringer04}
B{\"o}hringer, H., Schuecker, P., Guzzo, L., et al., 
2004, A\&A, 425, 367

\bibitem[2009]{bonafede09}
Bonafede, A., Feretti, L., Giovannini, G., et al.,
2009, A\&A, 503, 707

\bibitem[2007]{bondi07}
Bondi, M., Ciliegi, P., Venturi, T., et al.,
2007, A\&A, 463, 519 

\bibitem[2012]{boschin12}
Boschin, W., Girardi, M., Barrena, R., et al.,
2012, A\&A, 340A, 143

\bibitem[2012]{bourdin12}
Bourdin, H., Mazzotta, P., Markevitch, M., et al., 
2012, submitted to ApJ

\bibitem[2011]{shea11}
Brown, S., Rudnick, L.,
2011, MNRAS, 412, 2 

\bibitem[2001]{brunetti01}
Brunetti, G., Setti, G., Feretti, L., Giovannini, G., 
2001, MNRAS, 320, 365

\bibitem[2005]{bb05}
Brunetti, G., \& Blasi, P., 2005, MNRAS, 363, 1173

\bibitem[2007]{brunetti07}
Brunetti, G., Venturi, T., Dallacasa, D., et al.,
2007, ApJ, 670, L5

\bibitem[2008]{brunetti08}
Brunetti, G., Giacintucci, S., Cassano, R., et al., 
2008, Nature, 455, 944

\bibitem[2009]{brunetti09}
Brunetti, G., Cassano, R., Dolag, K., et al.,
2009, A\&A, 507, 661

\bibitem[2011]{bl11}
Brunetti, G., \& Lazarian, 2012, MNRAS, 412, 817

\bibitem[2011]{brunetti11}
Brunetti, G., 2011,
in {\it Non--thermal phenomena in colliding galaxy clusters},
Mem. SAIT, 82, 515

\bibitem[2001]{buote01}
Buote, D.A., 2001, ApJ 553, 15

\bibitem[2012]{carretti12}
Carretti, E., Brown, S., Stavaley--Smith, L., et al.,
2012, MNRAS, in press (arXiv:1205.1082v1)

\bibitem[2009]{cassano09}
Cassano, R., 2009, in {\it The low frequency radio Universe},
Eds. D.~J.~Saikia, D.~A.~Green, Y.~Gupta and T.~Venturi, 
ASP Conf. Ser. 407, 223

\bibitem[2005]{cassano05}
Cassano, R., Brunetti, G., 2005, MNRAS, 357, 1313

\bibitem[2006]{cassano06}
Cassano, R., Brunetti, G., Setti, G., 2006, 
MNRAS, 369, 1577 

\bibitem[2008]{cassano08}
Cassano, R., Brunetti, G., Venturi, T., et al.,  
2008, A\&A, 480, 687

\bibitem[2010]{cassano10}
Cassano, R., 2010, A\&A, 517A, 10

\bibitem[2010]{cassano10a}
Cassano, R., Ettori, S., Giacintucci, S., et al.,
2010, ApJ Letters, 721, L82

\bibitem[2011]{cassano11}
Cassano, R., Brunetti, G., Venturi, T.,
2011, Journal of Astroph. Astron., 32, 519

\bibitem[2006]{clarke06}
Clarke, T. E., \& En\ss{}lin, T. A., 2006, AJ, 131, 2900

\bibitem[2009]{dallacasa09}
Dallacasa, D., Brunetti, G., Giacintucci, S., et al.,
2009, ApJ, 699, 1288

\bibitem[2004]{DK04}
David, L.P., Kempner, J., 2004, ApJ, 613, 840

\bibitem[1980]{dennison1980}
Dennison, B., 1980, ApJ Letters, 239, L93

\bibitem[2011]{dwaraka11}
Dwarakanath, K.S., Rudnick, L., Shankar, N.U., Venturi, T., 2011,
{\it Diffuse Relativistic Plasmas}, Bangalore, 1--4 March 2010,
Conference Proceedings, Published by Indian Academy of Sciences, 
Journal of Astroph Astron., Vol. 32, N. 4 

\bibitem[1998]{ensslin98}
En\ss{}lin, T.A., Bierman, P.L., Klein, U., et al.,
1998, A\&A, 332 395

\bibitem[2001]{ensslin01}
En\ss{}lin, T., Gopal--Krishna, 2001, A\&A, 366, 26

\bibitem[2011]{ensslin11}
En\ss{}lin, T., Pfrommer, C., Miniati, F., et al.,
2011, A\&A, 527A, 99

\bibitem[2004]{feretti04}
Feretti, L., Brunetti, G., Giovannini, G., et al.,
2004, JKAS, 37,315

\bibitem[2005]{feretti05}
Feretti L., Schuecker, P., B\"oringer, H., et al.,
2005, A\&A, 444, 157

\bibitem[2012]{feretti12}
Feretti, L., Giovannini, G., Govoni, L., et al., 
2012, The Astronomy \& Astrophysics Review, 20, 54

\bibitem[2008]{ferrari08}
Ferrari, C., Govoni, F., Schindler, S., et al.,
2008, SSRv, 134, 93

\bibitem[2011]{ferrari11}
Ferrari, C., Br\"uggen, M., Brunetti, G., Venturi, T.,
2011, {\it Non--thermal phenomena in colliding galaxy clusters},
Nice 15--18 November 2010, Conference Proceedings, Editor F. Serra,
MemSAIt, Vol. 82, N. 3

\bibitem[2010]{finoguenov10}
Finoguenov, A., Sarazin, C. L., Nakazawa K., et al., 2010, ApJ, 715, 1143

\bibitem[2003]{fujita03}
Fujita, Y., Takizawa, M., \& Sarazin, C.L., 2003, ApJ, 584, 190

\bibitem[2005]{geller05}
Geller, M.J., Dell'Antonio, I.P., Kurtz, M.J., et al.,
2005, ApJ, 635L, 125

\bibitem[2008]{giacintucci08}
Giacintucci, S., Venturi, T., Macario, G., et al., 2008, A\&A, 486, 347

\bibitem[2009]{giacintucci09}
Giacintucci, S., Venturi, T., Brunetti, G., et al., 2009, A\&A, 505, 45

\bibitem[2013]{giacintucci13}
Giacintucci, S., Cassano, R., Brunetti, G., et al.,
2013, to be submitted to ApJ (G12)

\bibitem[2009]{giovannini09}
Giovannini, G., Bonafede, A., Feretti, L., et al.,
2010, A\&A, 507, 1257

\bibitem[2001]{GM01}
Girardi, M., Mezzetti, M., 2001, ApJ, 548, 79

\bibitem[2001]{govoni01}
Govoni, F., Feretti, L., Giovannini G., et al.,
2001, A\&A, 376, 803

\bibitem[2004]{govoni04}
Govoni, F., Markevitch, M., Vikhlinin A., et al.,
2004, ApJ, 605, 695

\bibitem[2011]{govoni11}
Govoni, F., Murgia, M., Giovannini, G., et al.,
2011, A\&A, 529A, 69

\bibitem[2010]{hao10}
Hao, J., McKay, T.A., Koester, B. et al., 
2010, ApJS, 191, 254

\bibitem[2007]{hoeft07}
Hoeft, M., Br\"uggen, M., 2007, MNRAS, 375, 77

\bibitem[1977]{jaffe77}
Jaffe, W.J., 1977, ApJ, 216,212

\bibitem[2001]{KS01}
Kempner, J.C., Sarazin, C.L., 2001, ApJ, 548, 639

\bibitem[2004]{KD04}
Kempner, J.C., David, L.P., 2004, MNRAS, 385, 392

\bibitem[2010]{keshet10}
Keshet, U., \& Loeb, A., 2010, ApJ, 722, 737

\bibitem[1989]{kim89}
Kim, K.-T., Kronberg, P.P., Giovannini, G., et al.,
1989, Nature, 341, 720

\bibitem[2007]{koester07}
Koester, B.P., McKay, T.A., Annis, J., et al.
2007, ApJ, 660, 239 

\bibitem[1997]{lemonon97}
Lemonon, L., Pierre, M., Hunstead, R.W., et al., 
1997, A\&A, 326, 34

\bibitem[2000]{liang00}
Liang, H., Hunstead, R.W., Birkinshaw, M., et al.,
2000, ApJ, 544, 686

\bibitem[2010]{macario10}
Macario, G., Venturi, T., Brunetti, G., et al., 
2010, A\&A, 517, A43

\bibitem[2011a]{macario11a}
Macario, G., Markevitch, M., Giacintucci, S., et al.,
2011a, ApJ, 728, 82

\bibitem[2011b]{macario11b}
Macario, G., Venturi. T., Dallacasa D., et al.,
2011b, in {\it Non--thermal phenomena in colliding galaxy clusters},
Mem. SAIT, 82, 557

\bibitem[2012]{macario12}
Macario, G., Venturi, T., Intema, H., et al., 2012,
submitted to A\&A

\bibitem[2002]{markevitch02}
Markevitch, M., Gonzalez, A. H., David, L., et al., 2002, ApJ, 567, 27

\bibitem[2005]{markevitch05}
Markevitch, M., Govoni, F., Brunetti, G., et al., 2005,
ApJ, 627, 733

\bibitem[2010]{markevitch10}
Markevitch, M., 2010, 12th Marcel Grossman Meeting, Paris,
arXiv:1010.3660v1

\bibitem[2011]{mazzotta11}
Mazzotta, P., Bourdin, H., Giacintucci, S., et al.,
2011, in {\it Non--thermal phenomena in colliding galaxy clusters},
Mem. SAIt, 82, 495

\bibitem[2011]{merten11}
Merten, J., Coe, D., Dupke, R., et al., 
2011, MNRAS, 417, 333

\bibitem[2003]{morrison03}
Morrison, G.E., {\bf et al.}, 2003, ApJSS, 146, 267

\bibitem[1985]{O'DO85}
O'Dea, C.P., Owen, F.N., 1985, AJ, 90, 954

\bibitem[1987]{odea87}
O'Dea, C.P., Sarazin, C.L., Owen, F.N., 1877, ApJ, 316, 113

\bibitem[2012]{ogrean12}
Ogrean, G., Br\"uggen, M., R\"ottgering, H., et al.,
2012, submitted to MNRAS (arXiv:1201.1502O)

\bibitem[2007]{orru07}
Orr\'u, E., Murgia, M., Feretti, L., et al.,
2007, A\&A, 467, 943

\bibitem[2012]{owers12}
Owers, M. S., Couch, W.J., Nulsen, P.E.J., et al., 
2012, ApJ, 750L, 23

\bibitem[2001]{petrosian01}
Petrosian, V., 2001, ApJ, 577, 560

\bibitem[2004]{pfrommer04}
Pfrommer, C., En\ss{}lin, T.A., 2004, JKAS, 37, 455

\bibitem[1997]{pierre97}
Pierre, M., Oukbir, J., Dubreuil, D., et al.,
1997, A\&AS, 124, 283

\bibitem[2009]{pizzo09}
Pizzo, R., de Bruyn, A.G., 2009, A\&A, 507, 639

\bibitem[2012]{ragozzine12}
Ragozzine, B., Clowe, D., Markevicth, M., et al.,
2012, ApJ, 744, 94

\bibitem[1999]{reid99}
Reid, A.D., Hunstead, R.W., Lemonon, L., et al.,
1999, MNRAS, 302,571 (R99)

\bibitem[1981]{roland81}
Roland, J., 1981, A\&A, 93, 407

\bibitem[2009]{rudnick09}
Rudnick, L., Lemmerman, J.A., 2009, 
ApJ, 697, 1341

\bibitem[1987]{schlickeiser87}
schlickeiser, R., Sievers, A., Thiemann, H., 1987,
A\&A, 182, 21

\bibitem[2002]{valtchanov02}
Valtchanov, I., Murphy, T., Pierre, M., et al., 2002,
A\&A, 392, 795

\bibitem[2010]{vanweeren10}
van Weeren, R.J., R\"{o}ttgering, H.J. A., Br\"{u}ggen, M., Hoeft, M.,
2010, Science, 330, 347

\bibitem[2011]{vanweeren11}
van Weeren, R.J., Br\"uggen, M., R\"ottgering, H.J.A., et al.,
2011, A\&A, 533, A35

\bibitem[2011b]{vanweeren11b}
van Weeren, R., Intema, H.T., R\"ottgering, H.J.A., et al.,
2011b, in {\it Non--thermal phenomena in colliding galaxy clusters},
MemSAIt, 75, 282

\bibitem[2012]{vanweeren12}
van Weeren, R.J., Bonafede, A., Ebeling, H., et al., 
2012, MNRAS, in press (arXiv:1206.2294)

\bibitem[2002]{venturi02}
Venturi, T., Bardelli, S., Zagaria, M., et al.,
2002, A\&A, 385, 39

\bibitem[2007]{venturi07}
Venturi, T., Giacintucci, S., Brunetti, G., et al., 2007, A\&A, 463, 937
(V07)

\bibitem[2008]{venturi08}
Venturi, T., Giacintucci, S., Dallacasa, D., et al., 2008, A\&A, 484, 327
(V08)

\bibitem[2011]{venturi11}
Venturi, T., 2011, in {\it Non--thermal phenomena in colliding galaxy clusters},
Mem. SAIT, 82, 499

\bibitem[2011a]{venturi11a}
Venturi, T., Giacintucci, S., Dallacasa, D., et al., 2011a, MNRAS Letters, 414, 
L65

\bibitem[2011b]{venturi11b}
Venturi, T., Giacintucci, S., Dallacasa, D., 2011b, Journal of Astroph.
and Astron., 32, 501

\bibitem[2012]{ziparo12}
Ziparo, F., Braglia, F.G., Pierini, D., et al.,
2012, MNRAS, 420, 2480

\end{thebibliography}
\end{document}